%% file: AA-2017-30746.tex
\documentclass[english]{aa}
\usepackage[T1]{fontenc}
\usepackage{verbatim}
\usepackage{amsbsy}
\usepackage{amstext}
\usepackage{graphicx}

\makeatletter
\usepackage{graphicx}
\usepackage[varg]{txfonts}

\makeatother

\usepackage{babel}
\begin{document}
\title{How much mass and angular momentum can the progenitors of carbon-enriched stars accrete?}

\author{E. Matrozis\thanks{Member of the International Max Planck Research School (IMPRS) for Astronomy and Astrophysics at the Universities of Bonn and Cologne}\and 
C. Abate \and  R. J. Stancliffe}

\institute{Argelander-Institut für Astronomie (AIfA), University of Bonn, Auf
dem Hügel 71, DE-53121, Bonn, Germany\\
\email{elvijs@astro.uni-bonn.de}}

\date{Received ; accepted}

\abstract{The chemically peculiar barium stars, CH stars, and most carbon-enhanced
metal-poor (CEMP) stars are all believed to be the products of mass
transfer in binary systems from a now extinct asymptotic giant branch
(AGB) primary star. The mass of the AGB star and the orbital parameters
of the system are the key factors usually considered when determining
how much mass is transferred onto the lower-mass main-sequence companion.
What is usually neglected, however, is the angular momentum of the
accreted material, which should spin up the accreting star. If the
star reaches critical rotation, further accretion should cease until
the excess angular momentum is somehow dealt with. If the star cannot
redistribute or lose the angular momentum while the primary is on
the AGB, the amount of mass accreted could be much lower than otherwise
expected. Here we present calculations, based on detailed stellar
evolution models, of the mass that can be accreted by putative progenitors
of Ba and CEMP stars before they reach critical rotation under the
assumption that no angular momentum loss occurs during the mass transfer.
We consider different accretion rates and values of specific angular
momentum. The most stringent limits on the accreted masses result
from considering accretion from a Keplerian accretion disk, which
is likely present during the formation of most extrinsically-polluted
carbon-enriched stars. Our calculations indicate that in this scenario
only about $0.05\ M_{\odot}$ of material can be added to the accreting
star before it reaches critical rotation, which is much too low to
explain the chemical enrichment of many Ba and CEMP stars. Either
the specific angular momentum of the accreted material has to effectively
be lower by about a factor of ten than the Keplerian value, or significant
angular momentum losses must occur for substantial accretion to take
place.}

\maketitle

\section{Introduction}

Barium stars, CH stars, and carbon-enhanced metal-poor (CEMP) stars
with \emph{s}-process enrichment (CEMP-\emph{s}) may seem rather distinct
objects. Barium stars are G- and K-type giants of roughly solar metallicity
with prominent lines of ionized barium and some molecular species
of carbon \citep{1951ApJ...114..473B}. CH stars are somewhat loosely
defined as stars whose spectra clearly show the presence of the CH
molecule in the photosphere \citep{1942ApJ....96..101K,1974ApJ...194...95B},
and are generally subgiants and giants with metallicity around $\text{[Fe/H]}=-1$.\footnotemark{}\footnotetext{The relative abundance of element A with respect to element B is $\text{[A/B]}=\log\left(C_\text{A}/C_\text{B}\right)-\log\left(C_\text{A}/C_\text{B}\right)_{\odot}$ where $C$ is the number or mass fraction.}
And CEMP-\emph{s} stars are low-metallicity ($\text{[Fe/H]}\lesssim-2$)
dwarfs and giants with substantial carbon and barium enrichment \citep[$\text{[C/Fe]}>1$, $\text{[Ba/Fe]}>1$, and $\text{[Ba/Eu]}>0.5$;][]{2005ARA&A..43..531B}.
Nevertheless, all of these objects share notable unifying characteristics:
while they display an enrichment in carbon and \emph{s}-process elements,
they are too unevolved to have produced these elements themselves;
their radial velocity variations indicate that they all host binary
companions \citep[but see][]{2016A&A...588A...3H}; the mass and luminosity
of the companion are consistent with that of a white dwarf \citep{1990ApJ...352..709M,2014MNRAS.441.1217S,2016A&A...586A.158J,2016A&A...586A.151M}.
These characteristics strongly suggest that all these carbon-enriched
objects are formed the same way: by binary mass transfer from an asymptotic
giant branch (AGB) companion that has since left the AGB and become
a white dwarf.

The amount of mass the progenitors of these carbon-enriched stars
can accrete must depend on the mass lost by the AGB star and the orbital
separation, which sets the mode of mass transfer between wind mass
transfer \citep{1995MNRAS.277.1443H,2003ASPC..303..290P}, wind Roche-lobe
overflow \citep[WRLOF;][]{2007ASPC..372..397M,2013A&A...552A..26A},
and Roche-lobe overflow (RLOF; likely of minor importance for these
systems as it should in most cases lead to common envelope evolution
and result in little accretion by the secondary). But it should also
depend on the angular momentum of the material raining down onto the
accretor. It seems plausible that, if the accreted material has enough
angular momentum to spin the accreting star up to critical rotation,
no further accretion can take place before the star either loses the
excess angular momentum, or somehow redistributes it in its interior.
\citet{1981A&A...102...17P} estimated analytically that a star needs
to accrete only about ten percent of its own mass before it reaches
critical rotation. However, the properties of many observed Ba and
CEMP-\emph{s} stars are hard to explain unless they have accreted
over 30\% of their initial mass \citep[e.g.][]{2013MNRAS.436.3068M,2015A&A...581A..22A,2015A&A...576A.118A}.

Here we investigate how much mass and angular momentum the supposed
progenitors of carbon-enriched stars can accrete before they reach
critical rotation. While we do not treat the physics of the accretion
process in detail, we consider different accretion rates and values
of the specific angular momentum of the accreted material. We thus
attempt to deduce if (and what amount of) angular momentum loss is
necessary to allow accreting enough mass to explain the chemical abundances
of observed CEMP-\emph{s} and Ba stars. While the focus is mainly
on the favourable case of instantaneous redistribution of angular
momentum throughout the star, such that uniform rotation is enforced
at all times, we also demonstrate the effect of a more realistic treatment
of angular momentum redistribution based on the diffusion approximation.

\section{Methods\label{sec:Methods}}

We use the stellar evolution code \textsc{stars} \citep{1971MNRAS.151..351E,1972MNRAS.156..361E,1995MNRAS.274..964P,2009MNRAS.396.1699S},
which has been extended to allow modelling of rotating stars by \citet{2012MNRAS.419..748P,2012MNRAS.423.1221P}
based on the work of \citet{1976ApJ...210..184E} and \citet{1997A&A...321..465M}.

Stellar rotation is inherently at least a two-dimensional problem.
For example, the local gravity $\boldsymbol{g}_{\text{eff}}$ in a
rotating star depends on the angle from the rotation axis $\theta$
because of the centrifugal force:
\begin{equation}
\boldsymbol{g}_{\text{eff}}=\left(-\frac{Gm}{r(\theta)^{2}}+\Omega^{2}r(\theta)\sin^{2}\theta\right)\boldsymbol{e}_{r}+\left(\Omega^{2}r(\theta)\sin\theta\cos\theta\right)\boldsymbol{e}_{\theta}.\label{eq:geff}
\end{equation}
(Here $\Omega$ is the angular velocity, $G$ is the gravitational
constant, and $m$ and $r$ are the mass and radial coordinate, respectively.)
However, the problem can be kept one-dimensional by formulating the
equations of stellar structure on isobaric surfaces $S_{P}$ characterized
by radius $r_{P}$, defined such that $V_{P}=\frac{4}{3}\pi r_{P}^{3}$
is the volume contained within $S_{P}$ \citep{1997A&A...321..465M}.
Quantities such as temperature and density then represent their averages
on an isobar. The radial coordinate $r_{P}$ corresponds to $r(\theta=\theta_{0}\approx54.7^{\circ}),$
where $\sin^{2}\theta_{0}=2/3$. When the star reaches critical rotation
($\Omega(\theta)=\Omega_{\text{c}}$), the radial component of the
effective gravity vanishes at the equator ($r(\pi/2)=r_{\text{e}}$).
At $\theta_{0}$ one has instead \citep{2009pfer.book.....M,2011A&A...527A..52G}
\begin{equation}
\frac{\frac{2}{3}\Omega_{\text{c}}^{2}r_{P,\text{c}}}{Gm/r_{P,\text{c}}^{2}}=\frac{2}{3}\left(\frac{r_{P,\text{c}}}{r_{\text{e,c}}}\right)^{3}\approx\frac{2}{3}\left(\frac{1.15}{1.5}\right)^{3}\approx0.3\label{eq:gcrit}
\end{equation}
or
\begin{equation}
\Omega_{\text{c}}\approx\sqrt{0.45\frac{Gm}{r_{P,\text{c}}^{3}}},\label{eq:omgc}
\end{equation}
which we adopt as the condition for critical rotation. This is close
to $\Omega_{\text{c}}=\sqrt{2Gm/(3r_{P,\text{c}}^{3})}$ used by \citet{2012MNRAS.419..748P},
to whom we refer for all the details behind the implementation of
rotation in the code.

Mass accretion is modelled by simply increasing the mass $M$ of a
model at a particular rate $\dot{M}$. The added mass is assigned
specific angular momentum $j_{\text{a}}$, so that addition of mass
$\Delta M$ results in the addition of total angular momentum $J_{\text{a}}=j_{\text{a}}\Delta M$.
To find the upper limit of $\Delta M$ for a given value of $j_{\text{a}}$,
we set the ZAMS rotation velocity to be small ($\Omega<0.01\Omega_{\text{c}}$),
so that the initial angular momentum is negligible. The mass is added
until $\Omega=\frac{3}{2}j/r^{2}$ reaches $\Omega_{\text{c}}$ anywhere
in the model (usually the surface). The added mass is set to have
the same composition as the surface at all times, i.e. we ignore composition
changes that would surely result from accretion of material from an
AGB star near the end of its life. This makes some difference to the
angular momentum that a star can accrete (Sect.~\ref{sec:Results}),
but none of the conclusions depend on this choice. Furthermore, we
assume that the entropy of the added matter is the same as that of
the surface at a given time, i.e. no additional energy is deposited
in the star by the in-falling matter. As a result of this assumption,
we probably underestimate the expansion that results from mass accretion
\citep{1985MNRAS.216...37P,1988AcA....38...89S,2010ApJ...721..478H,2016A&A...585A..65H}.
Relaxing this assumption would therefore revise our computed values
downwards.

The added angular momentum is instantaneously distributed throughout
the star so that uniform rotation results. This gives a plausible
upper limit to the amount of material with $j_{\text{a}}$ that a
star can accrete \citep{1981A&A...102...17P}. In addition, in some
models we follow the internal transport of angular momentum. \citet{2012MNRAS.419..748P}
discuss both advective \citep{1992A&A...265..115Z,1997A&A...317..749T}
and diffusive \citep{2000ApJ...528..368H} implementations. We opt
here for the diffusive approach, which the \textsc{stars} code is
more equipped to handle without additional calibrating constants.
Thus meridional circulation is approximated by the Eddington-Sweet
circulation \citep{1974IAUS...66...20K}, and the shear \citep{1974IAUS...59..185Z,1978ApJ...220..279E},
Solberg-H\o iland \citep{1946ApNr....4....1W}, and GSF \citep{1967ApJ...150..571G,1968ZA.....68..317F}
instabilities are all taken into account. For the parameters characterizing
the efficiency of rotational mixing, which here only play a minor
role because the composition of the accreted material is ignored,
we adopt $f_{\text{c}}=1/30$ and $f_{\mu}=0.05$, following \citet{2000ApJ...528..368H}.

Our models are characterized by the following set of parameters: metallicity
$Z$, initial primary mass $M_{\text{AGB}}$ (which here only sets
the age at the onset of mass transfer and limits the maximum mass
that can be accreted by the secondary to the total amount lost by
the primary), initial secondary mass $M$, mass accretion rate $\dot{M}$,
and specific angular momentum of accreted material $j_{\text{a}}$.
We restrict ourselves to $Z=10^{-4}$ and $M\simeq0.6\text{--}0.825~M_{\odot}$
for CEMP stars \citep{2015A&A...581A..22A,2015A&A...581A..62A}, and
$Z=0.008$ and $M\simeq1.0\text{--}2.5~M_{\odot}$ for Ba stars \citep[e.g.][]{1990ApJ...352..709M,2003ASPC..303..290P,2010A&A...523A..10I}.
The mass accretion rate and specific angular momentum are varied between
$\dot{M}\simeq10^{-8}\text{--}10^{-5}~M_{\odot}\text{yr}^{-1}$ and
$j_{\text{a}}=(0.2\text{--}3)\times10^{18}~\text{cm}^{2}\thinspace\text{s}^{-1}$.
The upper limit of $\dot{M}$ is close to typical mass-loss rates
of the donor stars near the end of their lives \citep{1993ApJ...413..641V,2005A&A...438..273V,2014A&A...566A.145R},
and the upper limit for $j_{\text{a}}$ is comparable to the Keplerian
value $j_{\text{K}}=\sqrt{GMR}\simeq(2\text{--}5)\times10^{18}\ \text{cm}^{2}\thinspace\text{s}^{-1}$
of the different progenitors. At all times we prevent $j_{\text{a}}$
from exceeding $j_{\text{K}}$.

\section{\label{sec:Results}Results}

We have computed the mass $\Delta M$ progenitors of carbon-enriched
stars can accrete prior to reaching critical rotation ($\Omega=\Omega_{\text{c}}$)
for a range of initial masses of the progenitor $M$, mass accretion
rates $\dot{M}$ and specific angular momenta of the added material
$j_{\text{a}}$. We now highlight the main features of the models
(a summary of all calculations is given in Table~\ref{tab:results}).

\begin{figure*}
\includegraphics[width=1\textwidth]{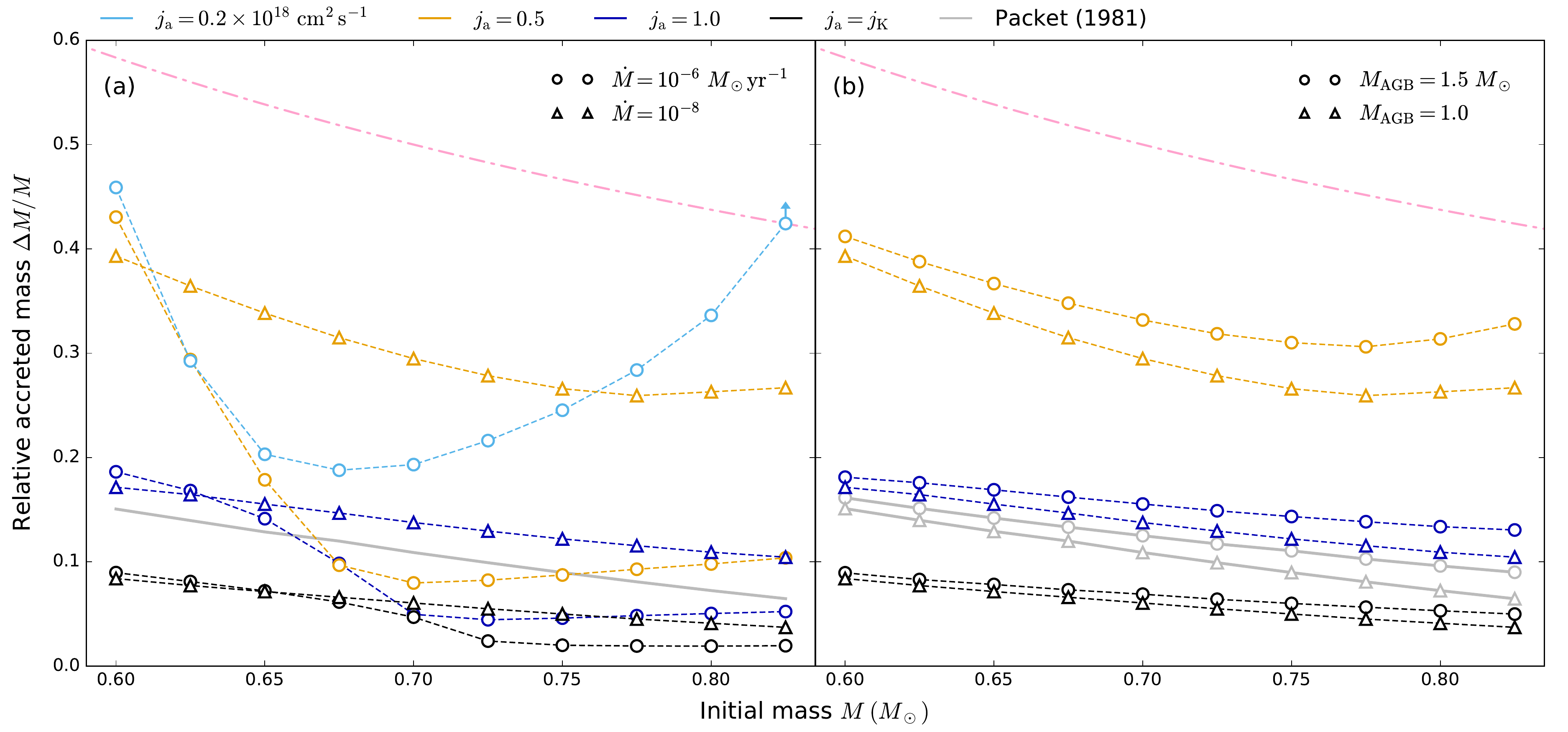}\caption{Mass accreted before critical rotation is reached (relative to initial
mass) by uniformly rotating CEMP-\emph{s} star progenitors for different
values of the specific angular momentum of accreted material $j_{\text{a}}$:
a) effect of different mass accretion rates in the $M_{\text{AGB}}=1\ M_{\odot}$
case); b) effect of different primary masses (age at the onset of
mass transfer) in the $\dot{M}=10^{-8}\ M_{\odot}\text{yr}^{-1}$
case. The case of $j_{\text{a}}=2\times10^{17}~\text{cm}^{2}\thinspace\text{s}^{-1}$,
$\dot{M}=10^{-8}~M_{\odot}\text{yr}^{-1}$ is not shown because in
all models $\Delta M$ exceeds $0.35\ M_{\odot}$ (dash-dotted line),
at which point the computations were stopped.\label{fig:z1e-4mp100s=000026eta1e_8s}}
\end{figure*}

Figure~\ref{fig:z1e-4mp100s=000026eta1e_8s} shows the amount of
mass CEMP-\emph{s} star progenitors of different masses can add before
reaching critical rotation. This amount is a complicated function
of the initial structure of the star, the angular momentum of the
added material, and the rate at which the material is added. Much
of this complexity is an outcome of the response of the star as it
gains mass. Thus one crucial parameter is the ratio between the mass
accretion timescale $\tau_{\dot{M}}=M/\dot{M}$ (here varied between
about $10^{5}$ and $10^{8}$ years) and the thermal adjustment (Kelvin-Helmholtz)
timescale $\tau_{\text{KH}}$ of the star (between about $10^{6}$
and close to $10^{8}$ years for the different progenitors). Generally
when $\tau_{\dot{M}}\gg\tau_{\text{KH}}$ the star is able to stay
close to thermal equilibrium, and its global properties change slowly.
The star evolves as if replaced at every instant by a slightly more
massive and more rapidly rotating star. In this case $\Delta M$ depends
almost entirely on the angular momentum of the added material alone
(this is more clearly seen in Fig.~\ref{fig:z1e-4mp100s=000026eta1e_8s}b,
where $\dot{M}=10^{-8}~M_{\odot}\text{yr}^{-1}$ and $\tau_{\dot{M}}>\tau_{\text{KH}}$
in all cases).

\begin{figure}
\includegraphics[width=1\columnwidth]{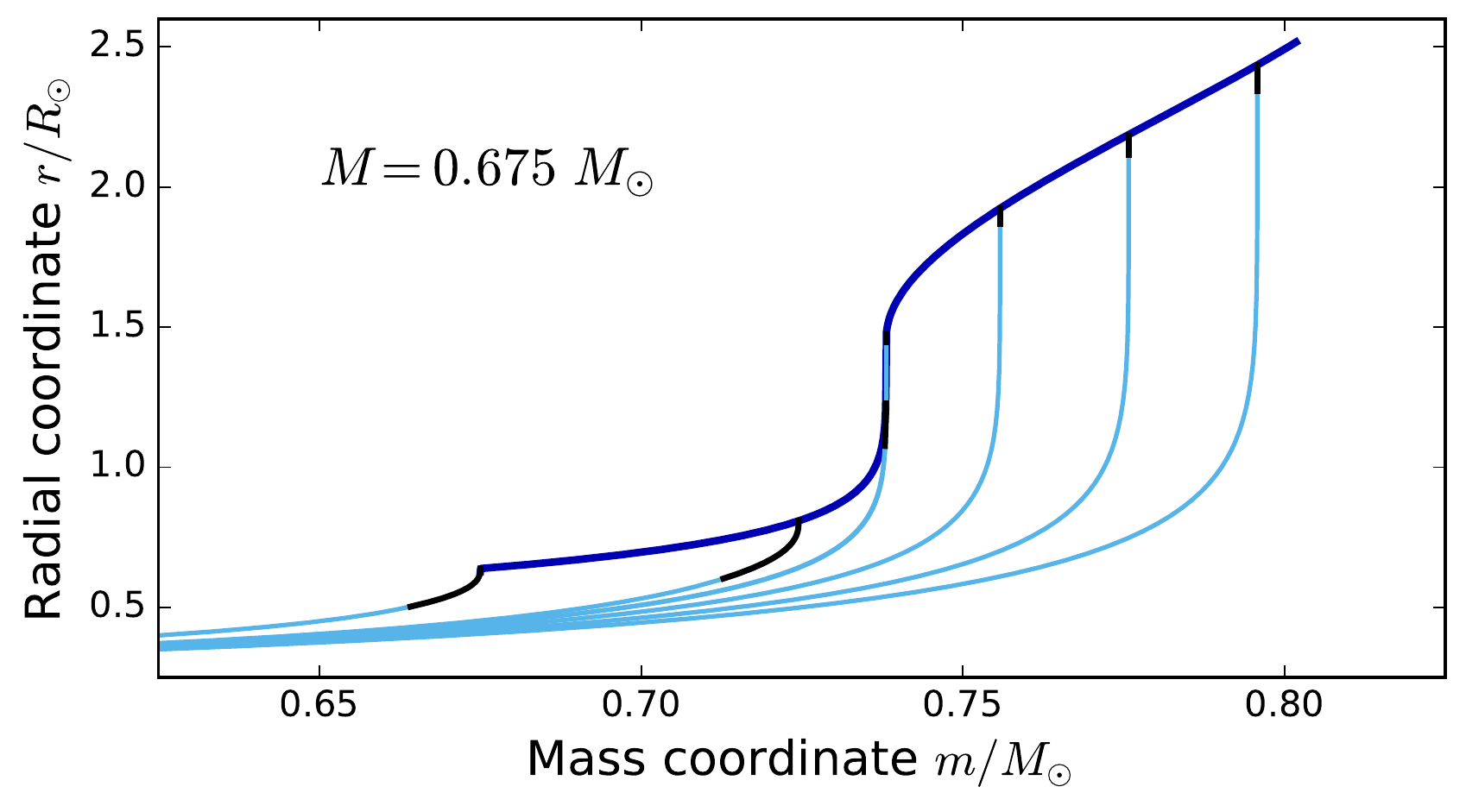}

\caption{Evolution of the mass distribution in a uniformly rotating model with
an initial mass of $0.675\ M_{\odot}$ when adding matter with $j_{\text{a}}=2\times10^{17}~\text{cm}^{2}\thinspace\text{s}^{-1}$
at a rate of $\dot{M}=10^{-6}\ M_{\odot}\text{yr}^{-1}$. The light-blue
lines show the radius variation with mass in six models from just
prior to mass addition to near critical rotation ($\Delta M\simeq0.125\ M_{\odot}$).
The black sections of the profiles indicate convective regions. The
upper (dark-blue) envelope shows the evolution of the surface radius
during mass addition.\label{fig:z1e-4mp100eta_1e-6j_2e+17_ms0675_r-vs-m}}
\end{figure}

\begin{figure}
\includegraphics[width=1\columnwidth]{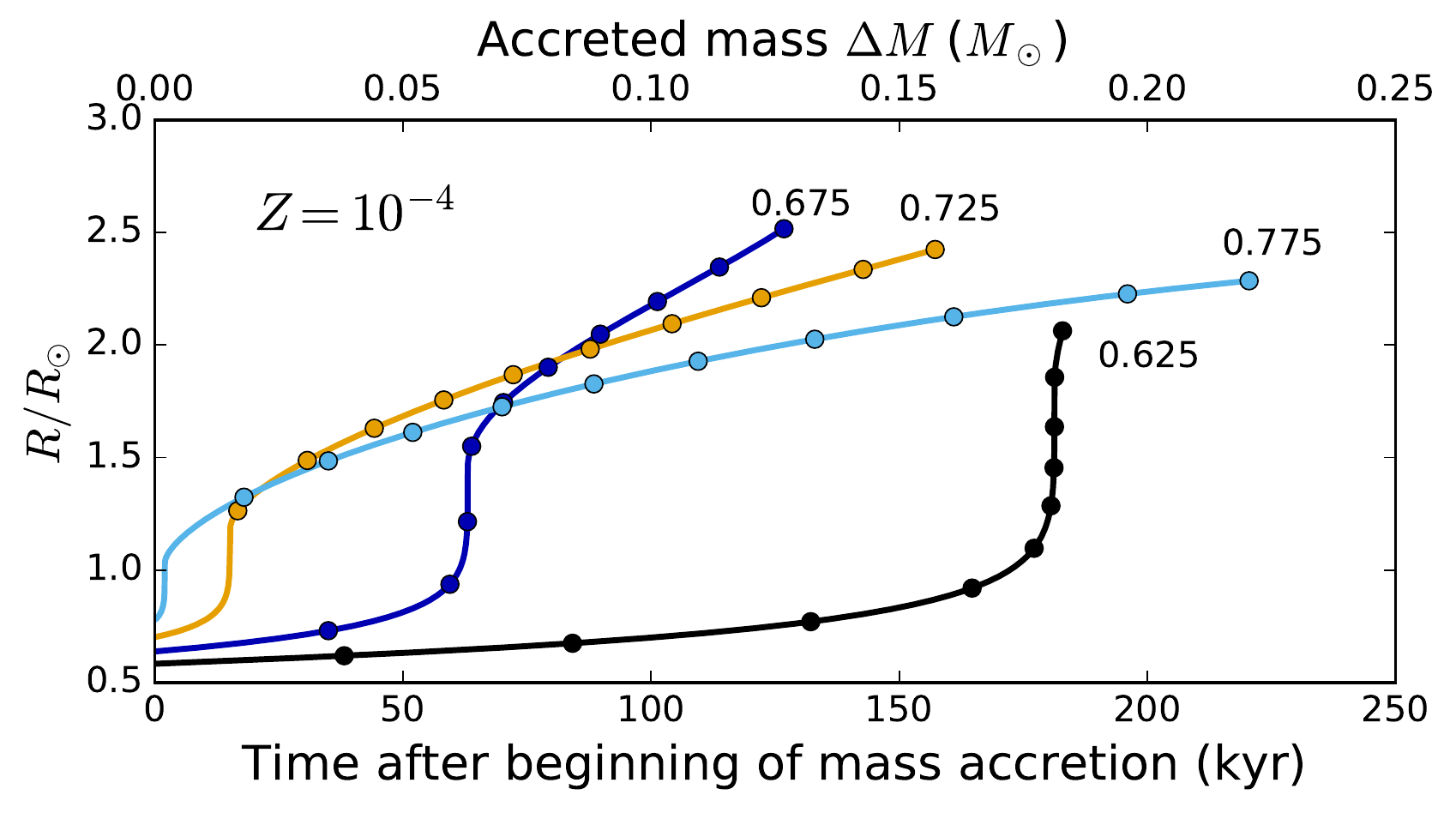}

\caption{Evolution of the radius in uniformly rotating models of the indicated
initial masses when adding matter with $j_{\text{a}}=2\times10^{17}~\text{cm}^{2}\thinspace\text{s}^{-1}$
at a rate of $\dot{M}=10^{-6}\ M_{\odot}\text{yr}^{-1}$. The markers
indicate every 10\% in $\Omega/\Omega_{\text{c}}$ with the last marker
indicating $\Omega=\Omega_{\text{c}}$. Prior to mass addition the
stars have a small outer convective region, which is more massive
in less massive stars. The disappearance of this region is followed
by a rapid expansion phase, after which the stars are closer to critical
rotation. The line corresponding to the $M=0.675\ M_{\odot}$ case
is the same as the upper envelope in Fig.~\ref{fig:z1e-4mp100eta_1e-6j_2e+17_ms0675_r-vs-m}.\label{fig:z1e-4mp100eta_1e-6j_2e+17_r-vs-t}}
\end{figure}

\begin{figure*}
\includegraphics[width=1\textwidth]{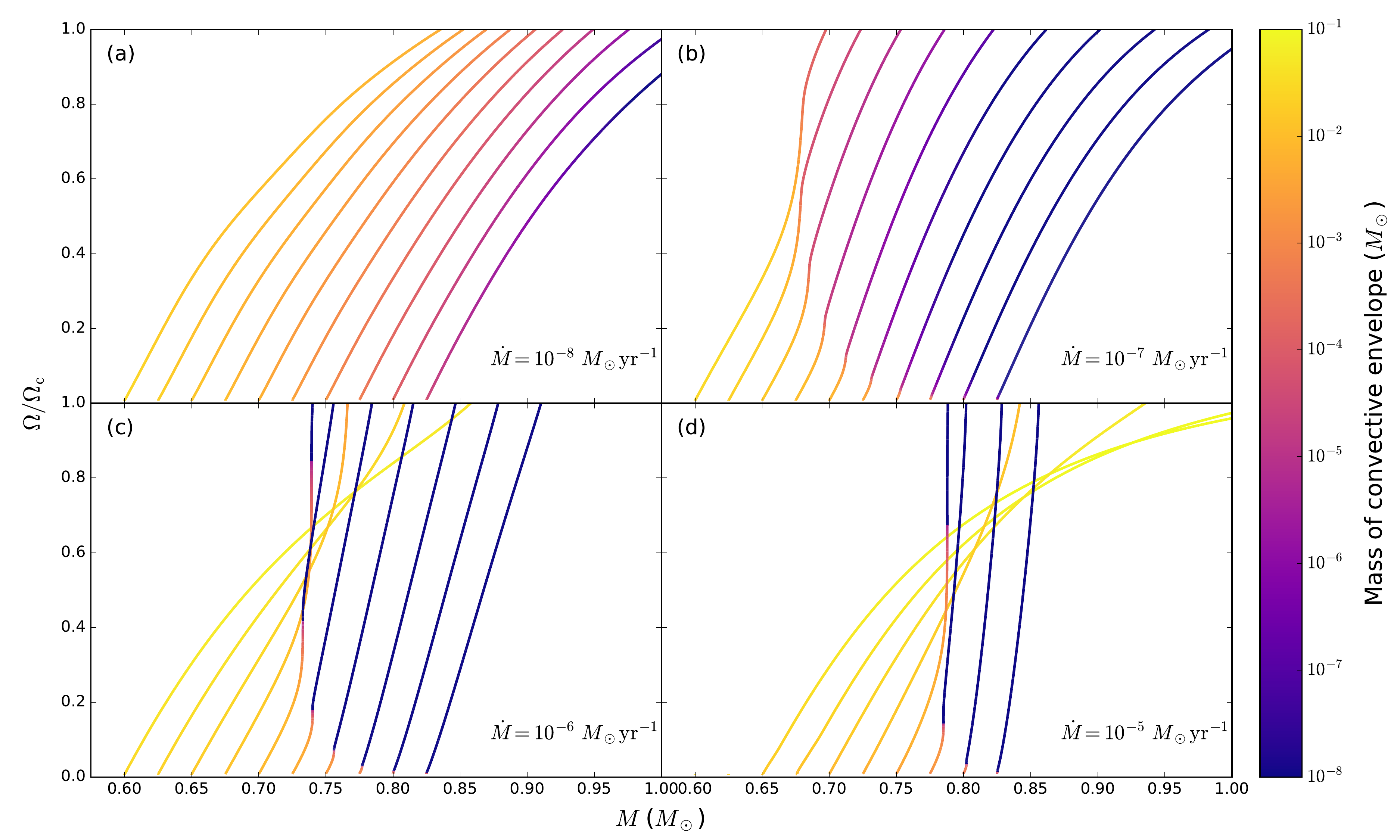}\caption{The evolution of the surface rotation velocity (starting from $\Omega/\Omega_{\text{c}}\approx0$)
for CEMP-\emph{s} star progenitors of different initial masses when
adding mass with $j_{\text{a}}=5\times10^{17}~\text{cm}^{2}\thinspace\text{s}^{-1}$.
The colour coding shows that the disappearance of the convective envelope
is accompanied by a rapid expansion phase, during which $\Omega/\Omega_{\text{c}}$
increases.\label{fig:z1e-4mp100j_5e+17s}}
\end{figure*}

When $\tau_{\dot{M}}\lesssim\tau_{\text{KH}}$, the star is driven
out of thermal equilibrium, forcing its outer layers to expand while
the inner regions are compressed. In this case seemingly minor differences
in the initial structure can lead to large differences in the response
of the star to mass addition. In particular, the response is related
to the convective stability of its outer layers. As long as there
is a substantial convective outer region, the star slowly expands
and heats up as mass is added. But if in the process of gaining mass
convection becomes inefficient and the convective region disappears,
the star experiences a brief phase of rapid expansion, during which
very little mass (and angular momentum) is accreted (Fig.~\ref{fig:z1e-4mp100eta_1e-6j_2e+17_ms0675_r-vs-m}).
As the star continues gaining mass, the expansion gradually slows
down and the surface layers begin to cool. In stars that initially
have a more massive convective region this phase plays a larger role
in bringing the star closer to critical rotation, but it also occurs
later (Figs.~\ref{fig:z1e-4mp100eta_1e-6j_2e+17_r-vs-t} and \ref{fig:z1e-4mp100j_5e+17s}).
The result can be a minimum in the $\Delta M(M)$ relation, as is
the case, e.g. at $M\simeq0.675\ M_{\odot}$ when $j_{\text{a}}=2\times10^{17}~\text{cm}^{2}\thinspace\text{s}^{-1}$
(Fig.~\ref{fig:z1e-4mp100s=000026eta1e_8s}a).

\begin{figure}
\includegraphics[width=1\columnwidth]{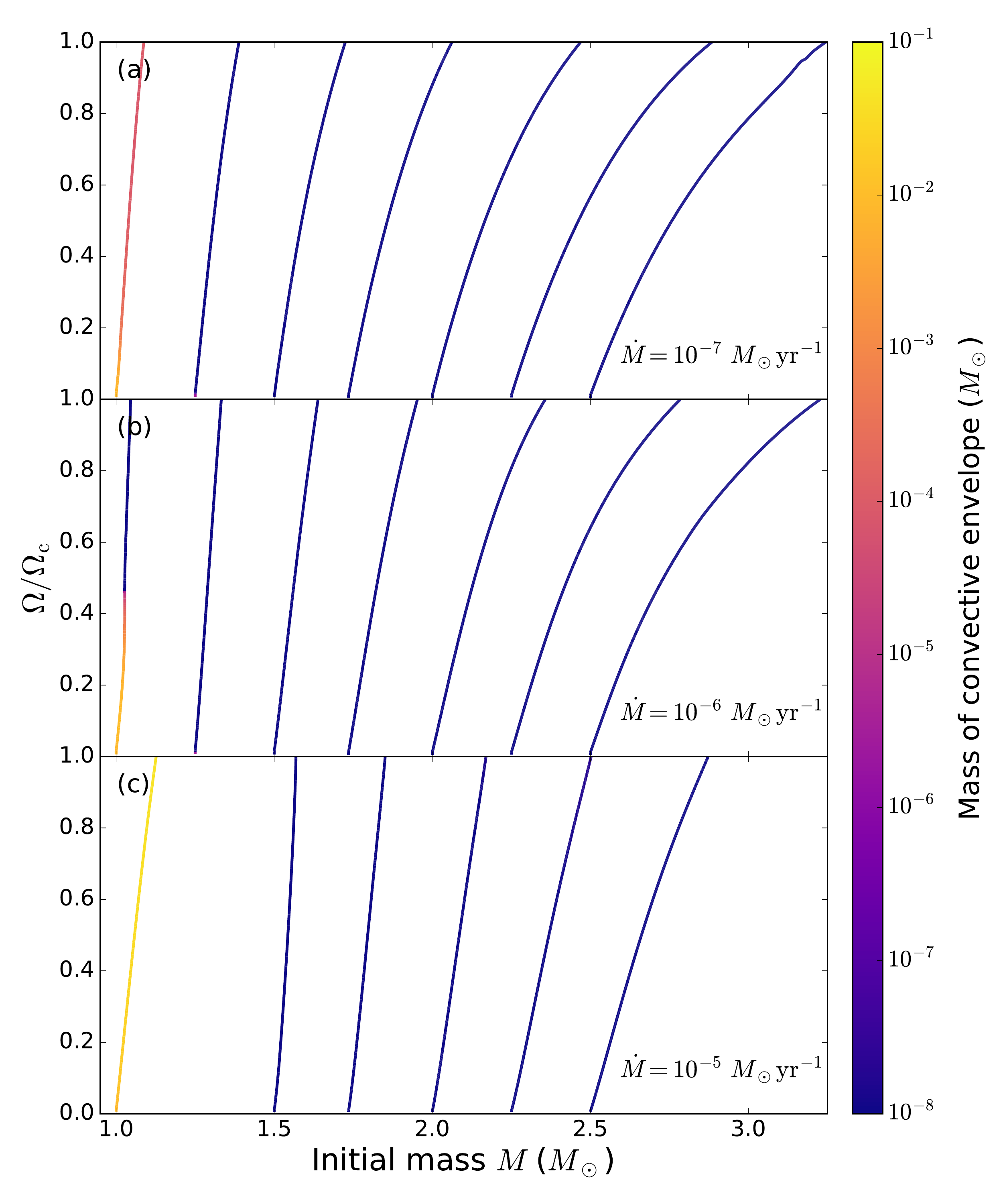}\caption{The evolution of the surface rotation velocity (starting from $\Omega/\Omega_{\text{c}}\approx0$)
for Ba star progenitors of different initial masses when adding mass
with $j_{\text{a}}=1\times10^{18}~\text{cm}^{2}\thinspace\text{s}^{-1}$.
Except for the $M=1\ M_{\odot}$ case, the outer regions of Ba star
progenitors remain radiative at all times.\label{fig:z8e-3mp300j1e+18s}}
\end{figure}

The evolution of accreting Ba star progenitors is often simpler. Owing
to their higher mass the thermal timescale of such stars is about
a factor of ten smaller than that of CEMP-\emph{s} star progenitors.
They are thus closer to thermal equilibrium during mass accretion.
Furthermore, except in the lowest mass cases ($M\simeq1\ M_{\odot}$),
these stars never have substantial convective envelopes. The disappearance
of the convective region and the rapid expansion phase, as observed
in the CEMP star case, therefore does not happen in most of these
models. Instead the critical rotation rate is reached more gradually,
and there is a smaller dependence on $\dot{M}$ (Fig.~\ref{fig:z8e-3mp300j1e+18s}).

\citet{1981A&A...102...17P} presents analytical estimates of the
amount of mass a star can accrete when $j_{\text{a}}=j_{\text{K}}$.
For an initially non-rotating star he finds approximately

\begin{equation}
\Delta M\simeq2\left(\sqrt{1+k^{2}}-1\right)M,\label{eq:dmp81}
\end{equation}
where $k^{2}=I/MR^{2}$, the square of the normalized gyration radius,
characterizes the distribution of mass inside a star with moment of
inertia $I$ (a smaller value of $k^{2}$ corresponds to a steeper,
more centrally concentrated mass distribution). While for CEMP-\emph{s}
star progenitors $k^{2}\simeq0.06\text{--}0.12$, for Ba star progenitors
$k^{2}\lesssim0.05$. Thus the value given by Eq.\ \eqref{eq:dmp81}
is similar for both sets of progenitors (approximately $0.07\text{--}0.09\ M_{\odot}$).
Our computed $\Delta M$ values in the $j_{\text{a}}=j_{\text{K}}$
case are also similar for the two sets, but they are about a factor
of two smaller than given by Eq.\ \eqref{eq:dmp81}. The difference
arises because the response of the star to mass addition (for example,
its change in $R$ and $k^{2}$) is not taken into account by \citet{1981A&A...102...17P}.
In particular, $k^{2}$ decreases during accretion when the outer
layers expand (Fig.~\ref{fig:z1e-4mp100eta_1e-6j_2e+17_ms0675_r-vs-m}),
which implies a lower value of $\Delta M$. Moreover, $\Delta M$
shows considerable variation with $\dot{M}$ (more so in the CEMP-\emph{s}
star case; Table\ \ref{tab:results}) because the response of the
star depends on the rate at which mass is added.

The distribution of mass in the progenitors of carbon-enriched stars
changes during their evolution such that $k^{2}$ decreases (the stars
become more centrally concentrated). As a consequence, the amount
of angular momentum the stars can accommodate also decreases over
time. Therefore, somewhat more material can be transferred in systems
hosting a more massive donor star in which mass transfer occurs earlier
(Fig.~\ref{fig:z1e-4mp100s=000026eta1e_8s}b). But note that we have
ignored the composition of the transferred material. Accretion of
material with a different composition will alter the opacity, and
hence the structure, of the outer layers. The composition will thus
also play a role in the response of the star to mass accretion and
the amount of mass needed to reach critical rotation. Test models
indicate that taking the composition into account can alter $\Delta M$
by some ten percent. The effect angular momentum accretion has on
the subsequent chemical evolution of carbon-enriched stars is explored
in \citet{Matrozis2017inpress}.

\begin{figure}
\includegraphics[width=1\columnwidth]{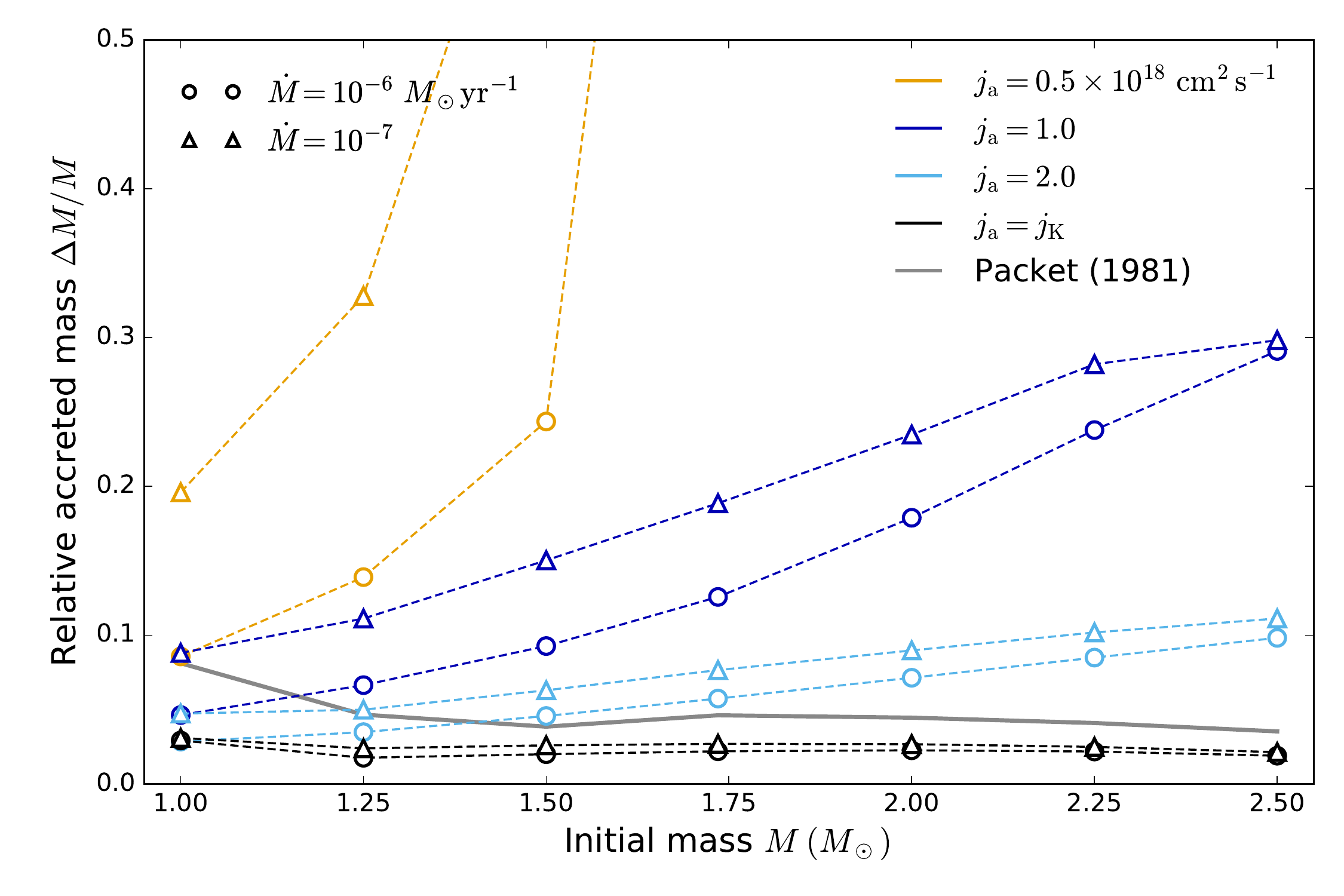}\caption{Mass accreted before critical rotation is reached (relative to initial
mass) by uniformly rotating Ba star progenitors for different values
of the specific angular momentum of accreted material. For progenitors
with $M\geq1.75\ M_{\odot}$ addition of material with $j_{\text{a}}\leq5\times10^{17}~\text{cm}^{2}\thinspace\text{s}^{-1}$
does not result in critical rotation even for $\Delta M>2\ M_{\odot}$.\label{fig:z8e-3mp300s}}
\end{figure}

What is the minimum specific angular momentum that material must have
such that accreting it can spin the star up to critical rotation?
More explicitly, what is $\langle j\rangle$ when going from $J_{0}=I_{0}\Omega_{0}=0$
to $J=I\Omega_{\text{c}}=\langle j\rangle\Delta M$ for large $\Delta M$?
Ignoring the change in $R$ and $k^{2}$ we can crudely estimate that
\begin{equation}
\langle j\rangle\simeq k^{2}R^{2}\left(1+\frac{M}{\Delta M}\right)\sqrt{G\left(M+\Delta M\right)/R^{3}}.\label{eq:jamin}
\end{equation}
For large $\Delta M$ (say $\Delta M\simeq M$) $\langle j\rangle\simeq3k^{2}j_{\text{K}}$,
which for typical values of $k^{2}$ gives $\langle j\rangle\gtrsim0.3j_{\text{K}}\simeq(5\text{--}7)\times10^{17}~\text{cm}^{2}\thinspace\text{s}^{-1}$
(CEMP stars) and $\langle j\rangle\gtrsim0.1j_{\text{K}}\simeq(5\text{--}8)\times10^{17}~\text{cm}^{2}\thinspace\text{s}^{-1}$
(Ba stars). These values are a bit high (Figs.~\ref{fig:z1e-4mp100s=000026eta1e_8s},
\ref{fig:z8e-3mp300s}), but give a reasonable estimate when the assumption
of a constant radius is closer to being satisfied (e.g. at low accretion
rates when thermal equilibrium is maintained). However, models that
do swell up can reach critical rotation for even lower values of $j_{\text{a}}$
(Fig.~\ref{fig:z1e-4mp100s=000026eta1e_8s}a).

\begin{figure}
\includegraphics[width=1\columnwidth]{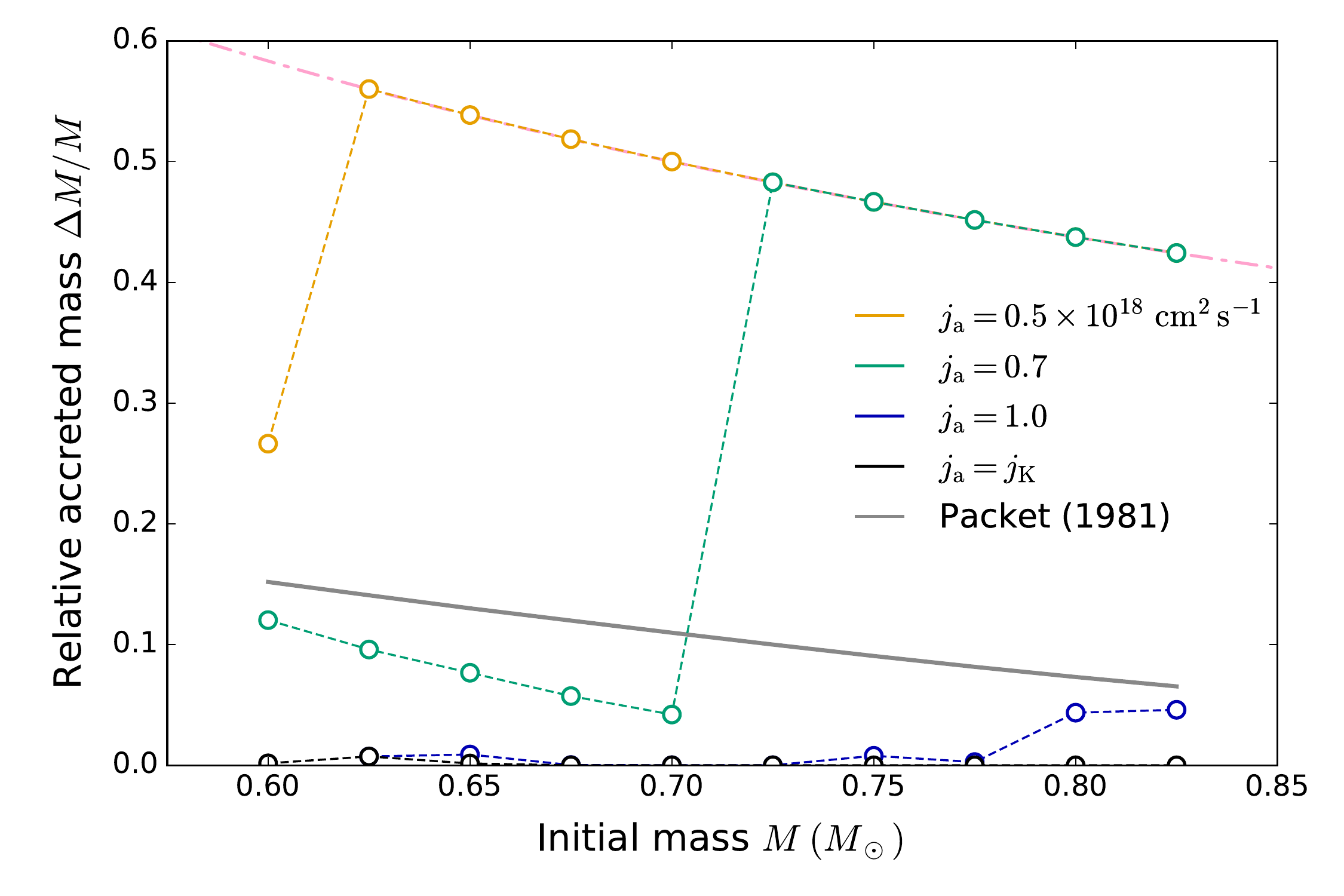}

\caption{Mass accreted before critical rotation is reached (relative to initial
mass) by differentially rotating CEMP-\emph{s} star progenitors for
different values of the specific angular momentum of accreted material.
Here $\dot{M}=10^{-6}\ M_{\odot}\text{yr}^{-1}$. The models were
stopped at $\Delta M=0.35\ M_{\odot}$ (dash-dotted line) if critical
rotation had not been reached by that point.\label{fig:z1e-4mp100eta1e-6d}}
\end{figure}

\begin{figure}
\includegraphics[width=1\columnwidth]{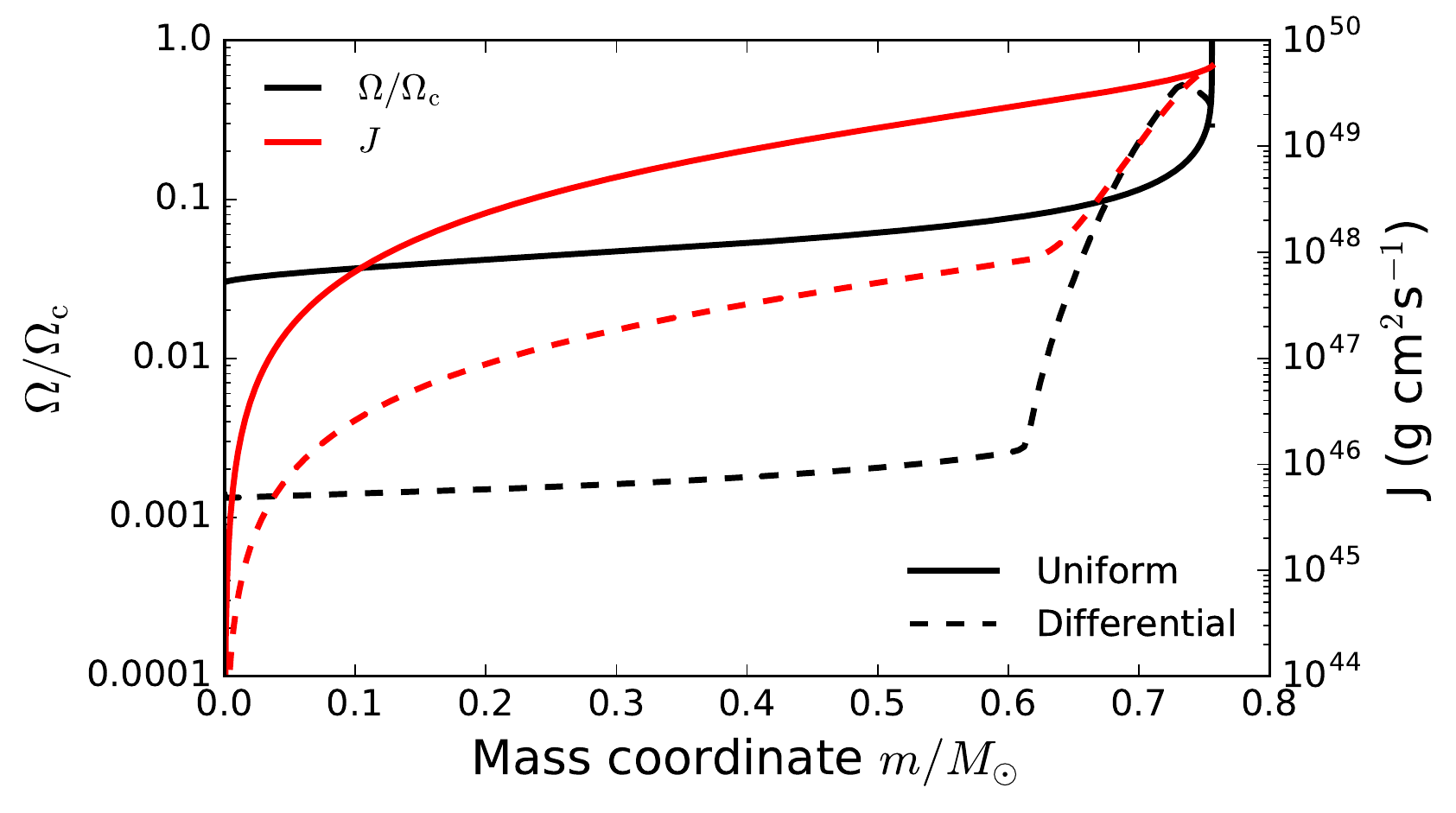}

\caption{The cumulative angular momentum and $\Omega/\Omega_{\text{c}}$ in
a $M=0.7\ M_{\odot}$ model that has added about $0.055\ M_{\odot}$
of matter with $j_{\text{a}}=5\times10^{17}~\text{cm}^{2}\thinspace\text{s}^{-1}$
at a rate of $\dot{M}=10^{-6}\ M_{\odot}\text{yr}^{-1}$. For the
uniformly rotating model this is just enough to bring it to critical
rotation, whereas in the differentially rotating model enough of the
angular momentum has been transported to layers deeper in the star
(between $m\simeq0.65$–$0.75\ M_{\odot}$) to prevent the surface
from rotating critically.\label{fig:z1e-4mp100ms0700eta_1e-6j_5e+17_d-vs-s}}
\end{figure}

When the redistribution of angular momentum is not instantaneous,
the angular momentum is rapidly distributed only in some outer part
of the star (e.g. the convective envelope, when present) instead of
throughout, and differential rotation results. One might then expect
that generally much less mass is necessary to spin the star up to
critical rotation. This is true for $j_{\text{a}}$ close to $j_{\text{K}}$
(having $j_{\text{a}}=j_{\text{K}}$ by definition sets the surface
to $\Omega=\Omega_{\text{c}}$). However, when $j_{\text{a}}$ is
substantially smaller than $j_{\text{K}}$, sometimes even more mass
can be added than in the uniformly rotating case (compare Figs.~\ref{fig:z1e-4mp100s=000026eta1e_8s}a
and \ref{fig:z1e-4mp100eta1e-6d}). For example, a $M=0.7\ M_{\odot}$
CEMP star progenitor rotating uniformly can only add about $0.05\ M_{\odot}$
of material (when $\dot{M}=10^{-6}\ M_{\odot}\text{yr}^{-1}$) with
$j_{\text{a}}=5\times10^{17}~\text{cm}^{2}\thinspace\text{s}^{-1}\simeq0.25j_{\text{K}}$
before reaching critical rotation. But in the differentially rotating
case even after adding $0.35\ M_{\odot}$ no part of the star has
reached $\Omega=\Omega_{\text{c}}$. Figure~\ref{fig:z1e-4mp100ms0700eta_1e-6j_5e+17_d-vs-s}
illustrates why this is the case. When $\Omega/\Omega_{\text{c}}$
reaches unity at the surface in the uniformly rotating case, most
of the interior of the star still has $\Omega/\Omega_{\text{c}}\ll1$
(solid black line). If more angular momentum could be stored in the
interior, the surface could be prevented from reaching $\Omega_{\text{c}}$.
This is what happens in the differentially rotating case where more
of the angular momentum is stored in the outer $0.1\ M_{\odot}$ of
material at the corresponding time (compare the red lines). By the
time $0.35\ M_{\odot}$ of material have been added ($0.35$~Myr
after the onset of mass accretion) the angular momentum has been transported
down to a mass coordinate of $m\simeq0.3\ M_{\odot}$, and $\Omega/\Omega_{\text{c}}>0.1$
in the outer $0.5\ M_{\odot}$. 

\section{Discussion\label{subsec:discussion}}

We have stopped the computations and recorded the accreted mass $\Delta M$
at the point when the model reaches critical rotation. But this $\Delta M$
is subject to change if critical rotation is reached when the star
is out of thermal equilibrium. Once accretion stops, the star will
attempt to regain equilibrium. Whether the star will tend to spin
up or down with respect to the critical rate as it evolves towards
equilibrium can be reasonably predicted by comparing $\Delta M$ with
$\Delta M_{\text{TE}}$, the value of $\Delta M$ in the case closest
to thermal equilibrium (case with the lowest $\dot{M}$).

Models with $\Delta M<\Delta M_{\text{TE}}$ usually experience a
phase in which very little mass and angular momentum are accreted
while $R$ and $\Omega/\Omega_{\text{c}}$ increase rapidly (Fig.\ \ref{fig:z1e-4mp100eta_1e-6j_2e+17_r-vs-t}).
After accretion these stars tend to contract and spin down with respect
to the critical rate. After moving away from critical rotation, they
could in principle go on to accrete more material (up to about $\Delta M_{\text{TE}}-\Delta M$
such that $\Delta M_{\text{final}}\simeq\Delta M_{\text{TE}}$). Conversely,
in cases where $\Delta M>\Delta M_{\text{TE}}$, the stars tend to
expand and spin up with respect to the critical rate. Since this is
not possible, they must somehow lose angular momentum to return to
equilibrium. This could happen by mass shedding \citep{2006A&A...447..623M,2011A&A...527A..84K}
so that some of the accreted material is, at least temporarily, lost.

The final $\Delta M$ thus depends on how the thermal relaxation timescale
compares to the accretion timescale. Given the large mass-loss rates
near the end of the AGB phase \citep[$\dot{M}_\text{AGB}\gtrsim 10^{-5}\,M_\odot\,\text{yr}^{-1}$;][]{2005A&A...438..273V,2014A&A...566A.145R},
the mass accretion rate in real systems is likely closer to the higher
range of values investigated here ($\dot{M}\simeq10^{-6\ldots-5}\ M_{\odot}\text{yr}^{-1}$,
depending on separation). Furthermore, this last stage, during which
the AGB star loses most of its envelope, lasts only a small fraction
of the overall thermally pulsing AGB lifetime of about 1\ Myr \citep{1993ApJ...413..641V,2007A&A...469..239M}
and is considerably shorter than the thermal timescales of the prospective
CEMP and Ba stars. It is thus difficult to envision how the progenitors
of carbon-enriched stars could maintain thermal equilibrium (and end
up with $\Delta M\simeq\Delta M_{\text{TE}}$) as they are accreting
mass from their AGB companion.

We have thus far not discussed the nature of mass transfer. Because
RLOF from an AGB star will in most cases be unstable, most carbon-enriched
stars probably form through wind mass transfer or WRLOF \citep{2007ASPC..372..397M,2013A&A...552A..26A}.
The specific angular momentum of accreted material in the canonical
Bondi-Hoyle accretion \citep{1944MNRAS.104..273B,2004NewAR..48..843E}
can be less than a few percent of $j_{\text{K}}$ \citep{2012ApJ...752...30B}
and would impose no limit on the accreted mass. But because of their
slow wind velocities, wind accretion from an AGB star is very different
from the Bondi-Hoyle case. As the slowly expanding material cools
and falls towards the accretor, it gains angular momentum from Coriolis
forces. Once the angular momentum exceeds the local Keplerian value
$\sqrt{GMr}$ ($r>R$), the material is expected to settle in a disk
around the accretor \citep{2006epbm.book.....E}. Both analytical
estimates \citep{2000ApJ...538..241S,2013ApJ...764..169P} and numerical
simulations \citep[e.g.][]{1996MNRAS.280.1264T,1998ApJ...497..303M,2009ApJ...700.1148D,2013MNRAS.433..295H,2017MNRAS.468.4465C}
suggest that disk formation in typical progenitor systems should be
common ($j_{\text{a}}\simeq j_{\text{K}}$).\footnote{Direct impact ($j_{\text{a}}<j_{\text{K}}$) of the transferring material
is unlikely because in all but the closest mass ratio binaries the
accreting star will be on the main sequence \citep{1991A&A...244..335N,2000ApJ...533..969B}
and thus well within its Roche lobe. This remains true even if the
stars swell up as a result of the accretion.} 

If the mass accretion in real systems indeed occurs at high rates
via a Keplerian disk, our results imply that most CEMP-\emph{s} (and
Ba) star progenitors can gain only up to about $\Delta M\simeq0.05\ M_{\odot}$
even if we assume that the angular momentum is rapidly distributed
throughout the star. This is a much smaller amount than the vast majority
of CEMP-\emph{s} stars in the population synthesis calculations of
\citet{2015A&A...581A..62A}, who find that the median mass accreted
is about $0.15\ M_{\odot}$ when angular momentum accretion is ignored.
Whether accounting for angular momentum accretion would present difficulties
explaining the observed fraction of CEMP stars \citep[more than ten percent of all stars with $\text{[Fe/H]}\lesssim -2$ are carbon-enhanced;][]{2006ApJ...652L..37L,2013AJ....146..132L,2014ApJ...797...21P}
would require a careful accounting of the number of systems expected
to be lost and gained as a result of severely limiting the accreted
masses. Most of the CEMP star progenitors in the simulations of \citet{2015A&A...581A..62A}
have initial masses well below $0.8\ M_{\odot}$. Stars with larger
initial masses ($M\simeq0.8\ M_{\odot}$) rapidly evolve after accreting
a large amount of mass and become white dwarfs. If such stars were
not able to gain as much mass, they might be around for long enough
to still be observable as CEMP stars. At the same time, some of the
lower mass stars, if they were to gain less mass, would no longer
be luminous enough to be observable \citep{2015A&A...581A..62A}.
At the moment we can only conclude that angular momentum considerations
suggest an initially more massive progenitor population of CEMP-\emph{s}
stars, compared to that found by \citet{2015A&A...581A..62A}.

A more serious issue is that often much higher accreted masses need
to be invoked to explain the properties of individual Ba and CEMP
stars. For example, \citet{2013MNRAS.436.3068M} require accreting
about $0.5\ M_{\odot}$ onto a $M\simeq1.5\ M_{\odot}$ (from a primary
with initial mass $M_{\text{AGB}}=1.8\ M_{\odot}$), or $0.8\ M_{\odot}$
onto a $M\simeq2\ M_{\odot}$ ($M_{\text{AGB}}=3\ M_{\odot}$) star
to explain the surface abundances of the Ba star in the planetary
nebula Hen 2-39%
. Many of the CEMP-\emph{s} stars considered by \citet{2015A&A...581A..22A,2015A&A...576A.118A}
are also best fit with accreted masses in excess of $0.1\ M_{\odot}$
given the current AGB nucleosynthesis model predictions \citep{2010MNRAS.403.1413K,2012ApJ...747....2L}.

In many cases the accreted masses needed to explain the surface chemistry
of carbon-enriched stars are a significant fraction of the total mass
lost by the AGB star, so the accretor must have captured nearly all
of the material entering its Roche lobe. If the star can somehow deal
with the angular momentum, this is not implausible. As the inner part
of the disk is accreted, the disk spreads outwards to compensate the
loss of angular momentum onto the accreting object \citep{1981ARA&A..19..137P,2008NewAR..52...21L}.
If a disk with initial radius $R_{\text{d}}$ spreads beyond the Roche
lobe radius $R_{\text{L}}$, the fraction of material lost that allows
the rest to fall to the central object is approximately $\sqrt{R_{\text{d}}/R_{\text{L}}}$
\citep{1976IAUS...73..237L}. For disks initially well inside the
Roche lobe this is a negligible amount. Furthermore, once the disk
spreads enough to approach $R_{\text{d}}\simeq R_{\text{L}}$, tides
from the AGB donor (or its remnant) should efficiently transfer angular
momentum from the outer disk edge back to the orbit, with little to
no mass lost from the Roche volume \citep{1977MNRAS.181..441P,1988AcA....38..189S,1994PASJ...46..621I,2005A&A...443..283H}.\footnote{The tidal influence on the accretor itself should be negligible since
the timescale for angular momentum transfer from the accretor back
to the orbit scales as $\tau_{\text{sync}}\sim(R/a)^{-6}$ \citep{1977A&A....57..383Z},
where $a\gg R$ is the orbital separation.}

If the star is spun up to critical rotation in the process, it needs
to lose angular momentum before it can accrete more of the disk. This
could be achieved by depositing the angular momentum back in the disk.
\citet{1991ApJ...370..604P} show that, once the star approaches critical
rotation, outward transport of angular momentum by shear stresses
($\dot{J}\propto\Sigma\nu\mathrm{d}{\Omega}/\mathrm{d}{r}$, where $\Sigma$ and
$\nu$ are respectively the surface density and viscosity of the disk)
can dominate the inward transport by accretion ($\dot{J}=\dot{M}j_{\text{a}}$),
and thus angular momentum can even be removed from the accreting object
as it gains mass, if there is sufficient viscous coupling between
the star and the disk. This coupling between the accreting star and
its disk could also be magnetic in nature. For example, in T~Tauri
stars (pre-main-sequence stars still surrounded by an accretion disk)
the magnetic field of the central star anchors in the disk, which
then exerts a negative torque on the star%
, preventing it from reaching critical rotation \citep[e.g. ][]{1996MNRAS.280..458A,1998ApJ...495..385H,2005MNRAS.356..167M}.

Also, some of the energy liberated during accretion could be used
to drive strong winds from the stellar surface. Ejecting a mere 10\%
of the accreted mass would suffice to remove enough angular momentum
to prevent critical rotation \citep{2005ApJ...632L.135M,2010MNRAS.406.1071D}.
However, surface magnetic field strengths in excess of $B\sim100\text{--}1000\ \text{G}$
are required for this mechanism or disk torques to be effective. While
such fields are probably uncommon in the progenitors of carbon-enriched
stars, they could perhaps be generated during the accretion if a strong
differential rotation results \citep{2010MNRAS.406.1071D}.

Whatever the mechanism, according to our calculations, for the accreted
masses to be large enough to satisfy the chemistry constraints (some
tenths of a solar mass), $j_{\text{a}}$ must effectively be below
about $5\times10^{17}~\text{cm}^{2}\thinspace\text{s}^{-1}$ (Figs.\ \ref{fig:z1e-4mp100s=000026eta1e_8s},
\ref{fig:z8e-3mp300s}), or about 10 and 25\% of $j_{\text{K}}$ for
Ba and CEMP stars, respectively. Put differently, above 75\% of the
angular momentum supplied to the accretor by the disk has to be removed
on average.

Nevertheless, the newly born carbon-enriched stars should rotate fairly
rapidly after the mass transfer ends. Indeed, a few younger Ba-enriched
stars with relatively rapid rotation velocities (tens of $\text{km}\,\text{s}^{-1}$
or more) are known \citep[e.g.][]{1995ApJ...438..364K,1996A&A...315L..19J,1998ApJ...502..763V,2003AJ....125..260B,2012MNRAS.419...39M},
indicating that some angular momentum accretion has occurred. Carbon-enhanced
metal-poor dwarfs also rotate more rapidly on average than other metal-poor
Halo stars \citep[about 10 and $3\,\text{km}\,\text{s}^{-1}$, respectively;][]{2012ApJ...751...14M,2003A&A...406..691L},
although the difference is small, which likely points to further angular
momentum loss following mass transfer. Given the large amount of time
elapsed since mass transfer took place (at least a gigayear)\footnote{This follows from comparing the lifetime (about 9 Gyr) of the lowest-mass
AGB star ($M_{\text{AGB}}\simeq0.9\ M_{\odot}$) that still undergoes
third dredge-up at $Z\simeq10^{-4}$ \citep{2008MNRAS.389.1828S,2010MNRAS.403.1413K}
with the youngest Halo stars (about 10 Gyr). At metallicities characteristic
of Ba stars there is no such lower limit.}, magnetic braking seems a plausible candidate for allowing the stars
to spin down, assuming it can work in stars with such small convective
envelopes \citep[$M_\text{env}<10^{-3}\,M_\odot$ following mass transfer;][]{2016A&A...592A..29M}.

\section{\label{sec:Conclusions}Conclusions}

We have calculated in an idealized way the amount of mass that can
be added to the progenitors of carbon-enriched (i.e. Ba and CEMP)
stars before they are spun up to critical rotation. Material assumed
to originate from a Keplerian accretion disk brings the stars to critical
rotation after only a few percent of their initial mass is added,
even if the angular momentum is allowed to rapidly redistribute throughout
the star. If instead the specific angular momentum of the accreted
material is a few tenths of the Keplerian value or less, the angular
momentum no longer limits the amount of mass that can be added.

Taken at face value, these results have major implications for the
progenitor systems of carbon-enriched stars, as many likely do accrete
matter through an accretion disk. The large accreted masses inferred
from observations of particular Ba and CEMP stars (in some instances
comparable to the initial mass of the progenitor) are at odds with
our results. However, a way to reconcile them with the necessity of
substantial mass accretion would be to invoke some mechanism, such
as viscous or magnetic coupling to the accretion disk, that would
allow the accreting star to transfer its spin angular momentum back
to the orbit with the help of tidal torques from the donor star. In
this way the accretor could possibly avoid critical rotation and accrete
much more of the material entering its Roche lobe.
\begin{acknowledgements}
We thank the referee Georges Meynet for valuable comments that have
helped improve the presentation and clarity of the paper. We thank
Adrian Potter for sharing with us his rotating version of the \textsc{stars}
code. We thank Rob Izzard and Zhengwei Liu for illuminating discussions.
CA acknowledges funding from the Alexander von Humboldt Foundation.
RJS is the recipient of a Sofja Kovalevskaja Award from the Alexander
von Humboldt Foundation.
\end{acknowledgements}


\include{tables/results_table}

\bibliographystyle{aa}

\bibliography{AA-2017-30746}
\end{document}

%% file: tables/results_table.tex
\begin{longtab}
\begin{longtable}{lllllllllllllllll}
\caption{Mass accreted before critical rotation is reached ($\Delta M$) by the different models.
The first five columns list the initial mass of the secondary ($M$);
its Kelvin-Helmholtz timescale before mass addition ($\tau_\text{KH}$, Myr);
its gyration radius before mass addition ($k^2=I/MR^{2}$);
estimate of $\Delta M$ based on \citet{1981A&A...102...17P} ($\Delta M_\text{P81}$);
mass accretion rate ($\dot{M}/10^{-6}\ M_{\odot}\text{yr}^{-1}$).
The remaining columns list pairs of
specific angular momentum of the accreted material ($j_\text{a}$ in units of $10^{18}\ \text{cm}^{2}\,\text{s}^{-1}$)
and the corresponding $\Delta M$.
The last pair in each row corresponds to $j_\text{a}=j_\text{K}$ and the listed value of $j_\text{a}$
is the Keplerian specific angular momentum at the surface prior to mass addition.
The table is sectioned according to the metallicity $Z$ and initial primary mass $M_\text{AGB}$ ($t_\text{mt}$ is the age at
which mass addition starts).
\label{tab:results}} \\
\hline
\hline
{\tiny{}$M$} & {\tiny{}$\tau_\text{KH}$} & {\tiny{}$k^2$} & {\tiny{}$\Delta M_\text{P81}$} & {\tiny{}$\dot{M}$} & {\tiny{}$j_\text{a}$} & {\tiny{}$\Delta M$} & {\tiny{}$j_\text{a}$} & {\tiny{}$\Delta M$} & {\tiny{}$j_\text{a}$} & {\tiny{}$\Delta M$} & {\tiny{}$j_\text{a}$} & {\tiny{}$\Delta M$} & {\tiny{}$j_\text{a}$} & {\tiny{}$\Delta M$} &{\tiny{}$j_\text{a}$} & {\tiny{}$\Delta M$} \tabularnewline
\hline
\endfirsthead
\caption{continued.}\\
\hline
\hline
{\tiny{}$M$} & {\tiny{}$\tau_\text{KH}$} & {\tiny{}$k^2$} & {\tiny{}$\Delta M_\text{P81}$} & {\tiny{}$\dot{M}$} & {\tiny{}$j_\text{a}$} & {\tiny{}$\Delta M$} & {\tiny{}$j_\text{a}$} & {\tiny{}$\Delta M$} & {\tiny{}$j_\text{a}$} & {\tiny{}$\Delta M$} & {\tiny{}$j_\text{a}$} & {\tiny{}$\Delta M$} & {\tiny{}$j_\text{a}$} & {\tiny{}$\Delta M$} & {\tiny{}$j_\text{a}$} & {\tiny{}$\Delta M$} \tabularnewline
\hline
\endhead
\hline
\endfoot
\multicolumn{17}{c}{{\tiny{}$Z=10^{-4}$; $M_\text{AGB}=1.0\ M_{\odot}$; $t_\mathrm{mt}=6.3\ \mathrm{Gyr}$; uniform rotation}}\tabularnewline
{\tiny{}$0.600$}  &  {\tiny{}$65.0$}  &  {\tiny{}$0.128$}  &  {\tiny{}$0.091$}  &  {\tiny{}$0.01$}  &  {\tiny{}$0.2$}  &  {\tiny{}$>0.35$}  &  {\tiny{}$0.5$}  &  {\tiny{}$0.236$}  &  {\tiny{}$1.0$}  &  {\tiny{}$0.103$}  &  {\tiny{}$2.0$}  &  {\tiny{}$0.050$}  &  {\tiny{}$1.758$}  &  {\tiny{}$0.050$}   & \ldots & \ldots  \tabularnewline
  \ldots  & \ldots & \ldots & \ldots  &  {\tiny{}$0.1$}  &  {\tiny{}$0.2$}  &  {\tiny{}$>0.35$}  &  {\tiny{}$0.5$}  &  {\tiny{}$0.098$}  &  {\tiny{}$1.0$}  &  {\tiny{}$0.079$}  &  {\tiny{}$2.0$}  &  {\tiny{}$0.050$}  &  {\tiny{}$1.758$}  &  {\tiny{}$0.050$}   & \ldots & \ldots  \tabularnewline
  \ldots  & \ldots & \ldots & \ldots  &  {\tiny{}$1.0$}  &  {\tiny{}$0.2$}  &  {\tiny{}$0.275$}  &  {\tiny{}$0.5$}  &  {\tiny{}$0.258$}  &  {\tiny{}$1.0$}  &  {\tiny{}$0.112$}  &  {\tiny{}$2.0$}  &  {\tiny{}$0.054$}  &  {\tiny{}$1.758$}  &  {\tiny{}$0.054$}   & \ldots & \ldots  \tabularnewline
  \ldots  & \ldots & \ldots & \ldots  &  {\tiny{}$10.0$}  &  {\tiny{}$0.2$}  &  {\tiny{}$  -  $}  &  {\tiny{}$0.5$}  &  {\tiny{}$  -  $}  &  {\tiny{}$1.0$}  &  {\tiny{}$  -  $}  &  {\tiny{}$2.0$}  &  {\tiny{}$  -  $}  &  {\tiny{}$1.758$}  &  {\tiny{}$  -  $}  & \ldots & \ldots  \tabularnewline
{\tiny{}$0.625$}  &  {\tiny{}$56.2$}  &  {\tiny{}$0.120$}  &  {\tiny{}$0.088$}  &  {\tiny{}$0.01$}  &  {\tiny{}$0.2$}  &  {\tiny{}$>0.35$}  &  {\tiny{}$0.5$}  &  {\tiny{}$0.228$}  &  {\tiny{}$1.0$}  &  {\tiny{}$0.103$}  &  {\tiny{}$2.0$}  &  {\tiny{}$0.049$}  &  {\tiny{}$1.836$}  &  {\tiny{}$0.048$}   & \ldots & \ldots  \tabularnewline
  \ldots  & \ldots & \ldots & \ldots  &  {\tiny{}$0.1$}  &  {\tiny{}$0.2$}  &  {\tiny{}$>0.35$}  &  {\tiny{}$0.5$}  &  {\tiny{}$0.099$}  &  {\tiny{}$1.0$}  &  {\tiny{}$0.060$}  &  {\tiny{}$2.0$}  &  {\tiny{}$0.045$}  &  {\tiny{}$1.836$}  &  {\tiny{}$0.044$}   & \ldots & \ldots  \tabularnewline
  \ldots  & \ldots & \ldots & \ldots  &  {\tiny{}$1.0$}  &  {\tiny{}$0.2$}  &  {\tiny{}$0.183$}  &  {\tiny{}$0.5$}  &  {\tiny{}$0.184$}  &  {\tiny{}$1.0$}  &  {\tiny{}$0.105$}  &  {\tiny{}$2.0$}  &  {\tiny{}$0.051$}  &  {\tiny{}$1.836$}  &  {\tiny{}$0.051$}   & \ldots & \ldots  \tabularnewline
  \ldots  & \ldots & \ldots & \ldots  &  {\tiny{}$10.0$}  &  {\tiny{}$0.2$}  &  {\tiny{}$  -  $}  &  {\tiny{}$0.5$}  &  {\tiny{}$  -  $}  &  {\tiny{}$1.0$}  &  {\tiny{}$  -  $}  &  {\tiny{}$2.0$}  &  {\tiny{}$  -  $}  &  {\tiny{}$1.836$}  &  {\tiny{}$  -  $}  & \ldots & \ldots  \tabularnewline
{\tiny{}$0.650$}  &  {\tiny{}$48.7$}  &  {\tiny{}$0.112$}  &  {\tiny{}$0.085$}  &  {\tiny{}$0.01$}  &  {\tiny{}$0.2$}  &  {\tiny{}$>0.35$}  &  {\tiny{}$0.5$}  &  {\tiny{}$0.220$}  &  {\tiny{}$1.0$}  &  {\tiny{}$0.101$}  &  {\tiny{}$2.0$}  &  {\tiny{}$0.047$}  &  {\tiny{}$1.914$}  &  {\tiny{}$0.046$}   & \ldots & \ldots  \tabularnewline
  \ldots  & \ldots & \ldots & \ldots  &  {\tiny{}$0.1$}  &  {\tiny{}$0.2$}  &  {\tiny{}$>0.35$}  &  {\tiny{}$0.5$}  &  {\tiny{}$0.104$}  &  {\tiny{}$1.0$}  &  {\tiny{}$0.052$}  &  {\tiny{}$2.0$}  &  {\tiny{}$0.038$}  &  {\tiny{}$1.914$}  &  {\tiny{}$0.037$}   & \ldots & \ldots \tabularnewline
  \ldots  & \ldots & \ldots & \ldots  &  {\tiny{}$1.0$}  &  {\tiny{}$0.2$}  &  {\tiny{}$0.132$}  &  {\tiny{}$0.5$}  &  {\tiny{}$0.116$}  &  {\tiny{}$1.0$}  &  {\tiny{}$0.092$}  &  {\tiny{}$2.0$}  &  {\tiny{}$0.048$}  &  {\tiny{}$1.914$}  &  {\tiny{}$0.047$}   & \ldots & \ldots \tabularnewline
  \ldots  & \ldots & \ldots & \ldots  &  {\tiny{}$10.0$}  &  {\tiny{}$0.2$}  &  {\tiny{}$  -  $}  &  {\tiny{}$0.5$}  &  {\tiny{}$>0.35$}  &  {\tiny{}$1.0$}  &  {\tiny{}$  -  $}  &  {\tiny{}$2.0$}  &  {\tiny{}$  -  $}  &  {\tiny{}$1.914$}  &  {\tiny{}$  -  $}  & \ldots & \ldots \tabularnewline
{\tiny{}$0.675$}  &  {\tiny{}$42.1$}  &  {\tiny{}$0.105$}  &  {\tiny{}$0.081$}  &  {\tiny{}$0.01$}  &  {\tiny{}$0.2$}  &  {\tiny{}$>0.35$}  &  {\tiny{}$0.5$}  &  {\tiny{}$0.213$}  &  {\tiny{}$1.0$}  &  {\tiny{}$0.099$}  &  {\tiny{}$2.0$}  &  {\tiny{}$0.047$}  &  {\tiny{}$1.995$}  &  {\tiny{}$0.045$}   & \ldots & \ldots  \tabularnewline
  \ldots  & \ldots & \ldots & \ldots  &  {\tiny{}$0.1$}  &  {\tiny{}$0.2$}  &  {\tiny{}$>0.35$}  &  {\tiny{}$0.5$}  &  {\tiny{}$0.111$}  &  {\tiny{}$1.0$}  &  {\tiny{}$0.052$}  &  {\tiny{}$2.0$}  &  {\tiny{}$0.029$}  &  {\tiny{}$1.995$}  &  {\tiny{}$0.028$}   & \ldots &  \ldots \tabularnewline
  \ldots  & \ldots & \ldots & \ldots  &  {\tiny{}$1.0$}  &  {\tiny{}$0.2$}  &  {\tiny{}$0.127$}  &  {\tiny{}$0.5$}  &  {\tiny{}$0.065$}  &  {\tiny{}$1.0$}  &  {\tiny{}$0.066$}  &  {\tiny{}$2.0$}  &  {\tiny{}$0.044$}  &  {\tiny{}$1.995$}  &  {\tiny{}$0.041$}   & \ldots &  \ldots \tabularnewline
  \ldots  & \ldots & \ldots & \ldots  &  {\tiny{}$10.0$}  &  {\tiny{}$0.2$}  &  {\tiny{}$>0.35$}  &  {\tiny{}$0.5$}  &  {\tiny{}$>0.35$}  &  {\tiny{}$1.0$}  &  {\tiny{}$0.109$}  &  {\tiny{}$2.0$}  &  {\tiny{}$0.048$}  &  {\tiny{}$1.995$}  &  {\tiny{}$0.046$}  & \ldots &  \ldots \tabularnewline
{\tiny{}$0.700$}  &  {\tiny{}$36.5$}  &  {\tiny{}$0.097$}  &  {\tiny{}$0.077$}  &  {\tiny{}$0.01$}  &  {\tiny{}$0.2$}  &  {\tiny{}$>0.35$}  &  {\tiny{}$0.5$}  &  {\tiny{}$0.206$}  &  {\tiny{}$1.0$}  &  {\tiny{}$0.096$}  &  {\tiny{}$2.0$}  &  {\tiny{}$0.047$}  &  {\tiny{}$2.080$}  &  {\tiny{}$0.042$}   & \ldots & \ldots  \tabularnewline
  \ldots  & \ldots & \ldots & \ldots  &  {\tiny{}$0.1$}  &  {\tiny{}$0.2$}  &  {\tiny{}$>0.35$}  &  {\tiny{}$0.5$}  &  {\tiny{}$0.123$}  &  {\tiny{}$1.0$}  &  {\tiny{}$0.054$}  &  {\tiny{}$2.0$}  &  {\tiny{}$0.028$}  &  {\tiny{}$2.080$}  &  {\tiny{}$0.024$}   & \ldots & \ldots  \tabularnewline
  \ldots  & \ldots & \ldots & \ldots  &  {\tiny{}$1.0$}  &  {\tiny{}$0.2$}  &  {\tiny{}$0.135$}  &  {\tiny{}$0.5$}  &  {\tiny{}$0.056$}  &  {\tiny{}$1.0$}  &  {\tiny{}$0.035$}  &  {\tiny{}$2.0$}  &  {\tiny{}$0.034$}  &  {\tiny{}$2.080$}  &  {\tiny{}$0.033$}   & \ldots & \ldots  \tabularnewline
  \ldots  & \ldots & \ldots & \ldots  &  {\tiny{}$10.0$}  &  {\tiny{}$0.2$}  &  {\tiny{}$0.297$}  &  {\tiny{}$0.5$}  &  {\tiny{}$0.236$}  &  {\tiny{}$1.0$}  &  {\tiny{}$0.099$}  &  {\tiny{}$2.0$}  &  {\tiny{}$0.046$}  &  {\tiny{}$2.080$}  &  {\tiny{}$0.042$}  & \ldots & \ldots  \tabularnewline
{\tiny{}$0.725$}  &  {\tiny{}$31.6$}  &  {\tiny{}$0.089$}  &  {\tiny{}$0.073$}  &  {\tiny{}$0.01$}  &  {\tiny{}$0.2$}  &  {\tiny{}$>0.35$}  &  {\tiny{}$0.5$}  &  {\tiny{}$0.202$}  &  {\tiny{}$1.0$}  &  {\tiny{}$0.094$}  &  {\tiny{}$2.0$}  &  {\tiny{}$0.046$}  &  {\tiny{}$2.168$}  &  {\tiny{}$0.040$}   & \ldots & \ldots  \tabularnewline
  \ldots  & \ldots & \ldots & \ldots  &  {\tiny{}$0.1$}  &  {\tiny{}$0.2$}  &  {\tiny{}$>0.35$}  &  {\tiny{}$0.5$}  &  {\tiny{}$0.137$}  &  {\tiny{}$1.0$}  &  {\tiny{}$0.056$}  &  {\tiny{}$2.0$}  &  {\tiny{}$0.029$}  &  {\tiny{}$2.168$}  &  {\tiny{}$0.024$}   & \ldots & \ldots  \tabularnewline
  \ldots  & \ldots & \ldots & \ldots  &  {\tiny{}$1.0$}  &  {\tiny{}$0.2$}  &  {\tiny{}$0.157$}  &  {\tiny{}$0.5$}  &  {\tiny{}$0.060$}  &  {\tiny{}$1.0$}  &  {\tiny{}$0.032$}  &  {\tiny{}$2.0$}  &  {\tiny{}$0.019$}  &  {\tiny{}$2.168$}  &  {\tiny{}$0.017$}   & \ldots & \ldots  \tabularnewline
  \ldots  & \ldots & \ldots & \ldots  &  {\tiny{}$10.0$}  &  {\tiny{}$0.2$}  &  {\tiny{}$0.118$}  &  {\tiny{}$0.5$}  &  {\tiny{}$0.117$}  &  {\tiny{}$1.0$}  &  {\tiny{}$0.079$}  &  {\tiny{}$2.0$}  &  {\tiny{}$0.042$}  &  {\tiny{}$2.168$}  &  {\tiny{}$0.036$}  & \ldots & \ldots  \tabularnewline
{\tiny{}$0.750$}  &  {\tiny{}$27.3$}  &  {\tiny{}$0.082$}  &  {\tiny{}$0.068$}  &  {\tiny{}$0.01$}  &  {\tiny{}$0.2$}  &  {\tiny{}$>0.35$}  &  {\tiny{}$0.5$}  &  {\tiny{}$0.199$}  &  {\tiny{}$1.0$}  &  {\tiny{}$0.091$}  &  {\tiny{}$2.0$}  &  {\tiny{}$0.045$}  &  {\tiny{}$2.262$}  &  {\tiny{}$0.037$}   & \ldots & \ldots  \tabularnewline
  \ldots  & \ldots & \ldots & \ldots  &  {\tiny{}$0.1$}  &  {\tiny{}$0.2$}  &  {\tiny{}$>0.35$}  &  {\tiny{}$0.5$}  &  {\tiny{}$0.152$}  &  {\tiny{}$1.0$}  &  {\tiny{}$0.060$}  &  {\tiny{}$2.0$}  &  {\tiny{}$0.030$}  &  {\tiny{}$2.262$}  &  {\tiny{}$0.023$}   & \ldots & \ldots  \tabularnewline
  \ldots  & \ldots & \ldots & \ldots  &  {\tiny{}$1.0$}  &  {\tiny{}$0.2$}  &  {\tiny{}$0.184$}  &  {\tiny{}$0.5$}  &  {\tiny{}$0.066$}  &  {\tiny{}$1.0$}  &  {\tiny{}$0.035$}  &  {\tiny{}$2.0$}  &  {\tiny{}$0.020$}  &  {\tiny{}$2.262$}  &  {\tiny{}$0.015$}   & \ldots & \ldots  \tabularnewline
  \ldots  & \ldots & \ldots & \ldots  &  {\tiny{}$10.0$}  &  {\tiny{}$0.2$}  &  {\tiny{}$0.051$}  &  {\tiny{}$0.5$}  &  {\tiny{}$0.037$}  &  {\tiny{}$1.0$}  &  {\tiny{}$0.039$}  &  {\tiny{}$2.0$}  &  {\tiny{}$0.032$}  &  {\tiny{}$2.262$}  &  {\tiny{}$0.028$}  & \ldots & \ldots  \tabularnewline
{\tiny{}$0.775$}  &  {\tiny{}$23.6$}  &  {\tiny{}$0.075$}  &  {\tiny{}$0.063$}  &  {\tiny{}$0.01$}  &  {\tiny{}$0.2$}  &  {\tiny{}$>0.35$}  &  {\tiny{}$0.5$}  &  {\tiny{}$0.201$}  &  {\tiny{}$1.0$}  &  {\tiny{}$0.089$}  &  {\tiny{}$2.0$}  &  {\tiny{}$0.043$}  &  {\tiny{}$2.361$}  &  {\tiny{}$0.035$}   & \ldots & \ldots  \tabularnewline
  \ldots  & \ldots & \ldots & \ldots  &  {\tiny{}$0.1$}  &  {\tiny{}$0.2$}  &  {\tiny{}$>0.35$}  &  {\tiny{}$0.5$}  &  {\tiny{}$0.168$}  &  {\tiny{}$1.0$}  &  {\tiny{}$0.064$}  &  {\tiny{}$2.0$}  &  {\tiny{}$0.030$}  &  {\tiny{}$2.361$}  &  {\tiny{}$0.023$}   & \ldots & \ldots  \tabularnewline
  \ldots  & \ldots & \ldots & \ldots  &  {\tiny{}$1.0$}  &  {\tiny{}$0.2$}  &  {\tiny{}$0.220$}  &  {\tiny{}$0.5$}  &  {\tiny{}$0.072$}  &  {\tiny{}$1.0$}  &  {\tiny{}$0.037$}  &  {\tiny{}$2.0$}  &  {\tiny{}$0.020$}  &  {\tiny{}$2.361$}  &  {\tiny{}$0.015$}   & \ldots & \ldots  \tabularnewline
  \ldots  & \ldots & \ldots & \ldots  &  {\tiny{}$10.0$}  &  {\tiny{}$0.2$}  &  {\tiny{}$0.048$}  &  {\tiny{}$0.5$}  &  {\tiny{}$0.027$}  &  {\tiny{}$1.0$}  &  {\tiny{}$0.018$}  &  {\tiny{}$2.0$}  &  {\tiny{}$0.012$}  &  {\tiny{}$2.361$}  &  {\tiny{}$0.011$}  & \ldots & \ldots  \tabularnewline
{\tiny{}$0.800$}  &  {\tiny{}$20.2$}  &  {\tiny{}$0.067$}  &  {\tiny{}$0.058$}  &  {\tiny{}$0.01$}  &  {\tiny{}$0.2$}  &  {\tiny{}$>0.35$}  &  {\tiny{}$0.5$}  &  {\tiny{}$0.210$}  &  {\tiny{}$1.0$}  &  {\tiny{}$0.087$}  &  {\tiny{}$2.0$}  &  {\tiny{}$0.042$}  &  {\tiny{}$2.466$}  &  {\tiny{}$0.033$}   & \ldots & \ldots  \tabularnewline
  \ldots  & \ldots & \ldots & \ldots  &  {\tiny{}$0.1$}  &  {\tiny{}$0.2$}  &  {\tiny{}$>0.35$}  &  {\tiny{}$0.5$}  &  {\tiny{}$0.183$}  &  {\tiny{}$1.0$}  &  {\tiny{}$0.068$}  &  {\tiny{}$2.0$}  &  {\tiny{}$0.032$}  &  {\tiny{}$2.466$}  &  {\tiny{}$0.023$}   & \ldots & \ldots  \tabularnewline
  \ldots  & \ldots & \ldots & \ldots  &  {\tiny{}$1.0$}  &  {\tiny{}$0.2$}  &  {\tiny{}$0.269$}  &  {\tiny{}$0.5$}  &  {\tiny{}$0.078$}  &  {\tiny{}$1.0$}  &  {\tiny{}$0.040$}  &  {\tiny{}$2.0$}  &  {\tiny{}$0.022$}  &  {\tiny{}$2.466$}  &  {\tiny{}$0.015$}   & \ldots & \ldots  \tabularnewline
  \ldots  & \ldots & \ldots & \ldots  &  {\tiny{}$10.0$}  &  {\tiny{}$0.2$}  &  {\tiny{}$0.053$}  &  {\tiny{}$0.5$}  &  {\tiny{}$0.029$}  &  {\tiny{}$1.0$}  &  {\tiny{}$0.019$}  &  {\tiny{}$2.0$}  &  {\tiny{}$0.011$}  &  {\tiny{}$2.466$}  &  {\tiny{}$0.008$}  & \ldots & \ldots  \tabularnewline
{\tiny{}$0.825$}  &  {\tiny{}$17.4$}  &  {\tiny{}$0.061$}  &  {\tiny{}$0.054$}  &  {\tiny{}$0.01$}  &  {\tiny{}$0.2$}  &  {\tiny{}$>0.35$}  &  {\tiny{}$0.5$}  &  {\tiny{}$0.220$}  &  {\tiny{}$1.0$}  &  {\tiny{}$0.086$}  &  {\tiny{}$2.0$}  &  {\tiny{}$0.041$}  &  {\tiny{}$2.578$}  &  {\tiny{}$0.031$}   & \ldots & \ldots  \tabularnewline
  \ldots  & \ldots & \ldots & \ldots  &  {\tiny{}$0.1$}  &  {\tiny{}$0.2$}  &  {\tiny{}$>0.35$}  &  {\tiny{}$0.5$}  &  {\tiny{}$0.196$}  &  {\tiny{}$1.0$}  &  {\tiny{}$0.072$}  &  {\tiny{}$2.0$}  &  {\tiny{}$0.034$}  &  {\tiny{}$2.578$}  &  {\tiny{}$0.024$}   & \ldots & \ldots  \tabularnewline
  \ldots  & \ldots & \ldots & \ldots  &  {\tiny{}$1.0$}  &  {\tiny{}$0.2$}  &  {\tiny{}$>0.35$}  &  {\tiny{}$0.5$}  &  {\tiny{}$0.086$}  &  {\tiny{}$1.0$}  &  {\tiny{}$0.043$}  &  {\tiny{}$2.0$}  &  {\tiny{}$0.023$}  &  {\tiny{}$2.578$}  &  {\tiny{}$0.016$}   & \ldots & \ldots  \tabularnewline
  \ldots  & \ldots & \ldots & \ldots  &  {\tiny{}$10.0$}  &  {\tiny{}$0.2$}  &  {\tiny{}$0.059$}  &  {\tiny{}$0.5$}  &  {\tiny{}$0.031$}  &  {\tiny{}$1.0$}  &  {\tiny{}$0.020$}  &  {\tiny{}$2.0$}  &  {\tiny{}$0.012$}  &  {\tiny{}$2.578$}  &  {\tiny{}$0.008$}  & \ldots & \ldots  \tabularnewline
\multicolumn{17}{c}{{\tiny{}$Z=10^{-4}$; $M_\text{AGB}=1.5\ M_{\odot}$; $t_\mathrm{mt}=1.8\ \mathrm{Gyr}$; uniform rotation}}\tabularnewline
{\tiny{}$0.600$}  &  {\tiny{}$72.2$}  &  {\tiny{}$0.135$}  &  {\tiny{}$0.098$}  &  {\tiny{}$0.01$}  &  {\tiny{}$0.2$}  &  {\tiny{}$>0.35$}  &  {\tiny{}$0.5$}  &  {\tiny{}$0.247$}  &  {\tiny{}$1.0$}  &  {\tiny{}$0.109$}  &  {\tiny{}$2.0$}  &  {\tiny{}$0.054$}  &  {\tiny{}$1.736$}  &  {\tiny{}$0.054$}   & \ldots & \ldots  \tabularnewline
  \ldots  & \ldots & \ldots & \ldots  &  {\tiny{}$0.1$}  &  {\tiny{}$0.2$}  &  {\tiny{}$>0.35$}  &  {\tiny{}$0.5$}  &  {\tiny{}$0.113$}  &  {\tiny{}$1.0$}  &  {\tiny{}$0.089$}  &  {\tiny{}$2.0$}  &  {\tiny{}$0.054$}  &  {\tiny{}$1.736$}  &  {\tiny{}$0.054$}   & \ldots & \ldots  \tabularnewline
  \ldots  & \ldots & \ldots & \ldots  &  {\tiny{}$1.0$}  &  {\tiny{}$0.2$}  &  {\tiny{}$  -  $}  &  {\tiny{}$0.5$}  &  {\tiny{}$0.330$}  &  {\tiny{}$1.0$}  &  {\tiny{}$0.120$}  &  {\tiny{}$2.0$}  &  {\tiny{}$0.058$}  &  {\tiny{}$1.736$}  &  {\tiny{}$0.058$}   & \ldots & \ldots  \tabularnewline
  \ldots  & \ldots & \ldots & \ldots  &  {\tiny{}$10.0$}  &  {\tiny{}$0.2$}  &  {\tiny{}$  -  $}  &  {\tiny{}$0.5$}  &  {\tiny{}$  -  $}  &  {\tiny{}$1.0$}  &  {\tiny{}$  -  $}  &  {\tiny{}$2.0$}  &  {\tiny{}$  -  $}  &  {\tiny{}$1.736$}  &  {\tiny{}$  -  $}  & \ldots & \ldots  \tabularnewline
{\tiny{}$0.625$}  &  {\tiny{}$63.5$}  &  {\tiny{}$0.128$}  &  {\tiny{}$0.095$}  &  {\tiny{}$0.01$}  &  {\tiny{}$0.2$}  &  {\tiny{}$>0.35$}  &  {\tiny{}$0.5$}  &  {\tiny{}$0.242$}  &  {\tiny{}$1.0$}  &  {\tiny{}$0.110$}  &  {\tiny{}$2.0$}  &  {\tiny{}$0.052$}  &  {\tiny{}$1.809$}  &  {\tiny{}$0.052$}   & \ldots & \ldots  \tabularnewline
  \ldots  & \ldots & \ldots & \ldots  &  {\tiny{}$0.1$}  &  {\tiny{}$0.2$}  &  {\tiny{}$>0.35$}  &  {\tiny{}$0.5$}  &  {\tiny{}$0.104$}  &  {\tiny{}$1.0$}  &  {\tiny{}$0.072$}  &  {\tiny{}$2.0$}  &  {\tiny{}$0.050$}  &  {\tiny{}$1.809$}  &  {\tiny{}$0.050$}   & \ldots & \ldots  \tabularnewline
  \ldots  & \ldots & \ldots & \ldots  &  {\tiny{}$1.0$}  &  {\tiny{}$0.2$}  &  {\tiny{}$0.229$}  &  {\tiny{}$0.5$}  &  {\tiny{}$0.228$}  &  {\tiny{}$1.0$}  &  {\tiny{}$0.116$}  &  {\tiny{}$2.0$}  &  {\tiny{}$0.055$}  &  {\tiny{}$1.809$}  &  {\tiny{}$0.055$}   & \ldots & \ldots  \tabularnewline
  \ldots  & \ldots & \ldots & \ldots  &  {\tiny{}$10.0$}  &  {\tiny{}$0.2$}  &  {\tiny{}$  -  $}  &  {\tiny{}$0.5$}  &  {\tiny{}$  -  $}  &  {\tiny{}$1.0$}  &  {\tiny{}$  -  $}  &  {\tiny{}$2.0$}  &  {\tiny{}$  -  $}  &  {\tiny{}$1.809$}  &  {\tiny{}$  -  $}  & \ldots & \ldots  \tabularnewline
{\tiny{}$0.650$}  &  {\tiny{}$56.1$}  &  {\tiny{}$0.122$}  &  {\tiny{}$0.093$}  &  {\tiny{}$0.01$}  &  {\tiny{}$0.2$}  &  {\tiny{}$>0.35$}  &  {\tiny{}$0.5$}  &  {\tiny{}$0.238$}  &  {\tiny{}$1.0$}  &  {\tiny{}$0.110$}  &  {\tiny{}$2.0$}  &  {\tiny{}$0.051$}  &  {\tiny{}$1.881$}  &  {\tiny{}$0.051$}   & \ldots & \ldots  \tabularnewline
  \ldots  & \ldots & \ldots & \ldots  &  {\tiny{}$0.1$}  &  {\tiny{}$0.2$}  &  {\tiny{}$>0.35$}  &  {\tiny{}$0.5$}  &  {\tiny{}$0.108$}  &  {\tiny{}$1.0$}  &  {\tiny{}$0.055$}  &  {\tiny{}$2.0$}  &  {\tiny{}$0.045$}  &  {\tiny{}$1.881$}  &  {\tiny{}$0.044$}   & \ldots & \ldots  \tabularnewline
  \ldots  & \ldots & \ldots & \ldots  &  {\tiny{}$1.0$}  &  {\tiny{}$0.2$}  &  {\tiny{}$0.164$}  &  {\tiny{}$0.5$}  &  {\tiny{}$0.160$}  &  {\tiny{}$1.0$}  &  {\tiny{}$0.108$}  &  {\tiny{}$2.0$}  &  {\tiny{}$0.053$}  &  {\tiny{}$1.881$}  &  {\tiny{}$0.052$}   & \ldots & \ldots  \tabularnewline
  \ldots  & \ldots & \ldots & \ldots  &  {\tiny{}$10.0$}  &  {\tiny{}$0.2$}  &  {\tiny{}$  -  $}  &  {\tiny{}$0.5$}  &  {\tiny{}$  -  $}  &  {\tiny{}$1.0$}  &  {\tiny{}$  -  $}  &  {\tiny{}$2.0$}  &  {\tiny{}$  -  $}  &  {\tiny{}$1.881$}  &  {\tiny{}$  -  $}  & \ldots & \ldots  \tabularnewline
{\tiny{}$0.675$}  &  {\tiny{}$49.6$}  &  {\tiny{}$0.115$}  &  {\tiny{}$0.091$}  &  {\tiny{}$0.01$}  &  {\tiny{}$0.2$}  &  {\tiny{}$>0.35$}  &  {\tiny{}$0.5$}  &  {\tiny{}$0.235$}  &  {\tiny{}$1.0$}  &  {\tiny{}$0.109$}  &  {\tiny{}$2.0$}  &  {\tiny{}$0.051$}  &  {\tiny{}$1.954$}  &  {\tiny{}$0.049$}   & \ldots & \ldots  \tabularnewline
  \ldots  & \ldots & \ldots & \ldots  &  {\tiny{}$0.1$}  &  {\tiny{}$0.2$}  &  {\tiny{}$>0.35$}  &  {\tiny{}$0.5$}  &  {\tiny{}$0.117$}  &  {\tiny{}$1.0$}  &  {\tiny{}$0.055$}  &  {\tiny{}$2.0$}  &  {\tiny{}$0.037$}  &  {\tiny{}$1.954$}  &  {\tiny{}$0.036$}   & \ldots & \ldots  \tabularnewline
  \ldots  & \ldots & \ldots & \ldots  &  {\tiny{}$1.0$}  &  {\tiny{}$0.2$}  &  {\tiny{}$0.142$}  &  {\tiny{}$0.5$}  &  {\tiny{}$0.102$}  &  {\tiny{}$1.0$}  &  {\tiny{}$0.092$}  &  {\tiny{}$2.0$}  &  {\tiny{}$0.051$}  &  {\tiny{}$1.954$}  &  {\tiny{}$0.049$}   & \ldots & \ldots  \tabularnewline
  \ldots  & \ldots & \ldots & \ldots  &  {\tiny{}$10.0$}  &  {\tiny{}$0.2$}  &  {\tiny{}$  -  $}  &  {\tiny{}$0.5$}  &  {\tiny{}$  -  $}  &  {\tiny{}$1.0$}  &  {\tiny{}$  -  $}  &  {\tiny{}$2.0$}  &  {\tiny{}$  -  $}  &  {\tiny{}$1.954$}  &  {\tiny{}$  -  $}  & \ldots & \ldots  \tabularnewline
{\tiny{}$0.700$}  &  {\tiny{}$44.1$}  &  {\tiny{}$0.109$}  &  {\tiny{}$0.088$}  &  {\tiny{}$0.01$}  &  {\tiny{}$0.2$}  &  {\tiny{}$>0.35$}  &  {\tiny{}$0.5$}  &  {\tiny{}$0.232$}  &  {\tiny{}$1.0$}  &  {\tiny{}$0.109$}  &  {\tiny{}$2.0$}  &  {\tiny{}$0.052$}  &  {\tiny{}$2.028$}  &  {\tiny{}$0.048$}   & \ldots & \ldots  \tabularnewline
  \ldots  & \ldots & \ldots & \ldots  &  {\tiny{}$0.1$}  &  {\tiny{}$0.2$}  &  {\tiny{}$>0.35$}  &  {\tiny{}$0.5$}  &  {\tiny{}$0.130$}  &  {\tiny{}$1.0$}  &  {\tiny{}$0.057$}  &  {\tiny{}$2.0$}  &  {\tiny{}$0.031$}  &  {\tiny{}$2.028$}  &  {\tiny{}$0.028$}   & \ldots & \ldots  \tabularnewline
  \ldots  & \ldots & \ldots & \ldots  &  {\tiny{}$1.0$}  &  {\tiny{}$0.2$}  &  {\tiny{}$0.152$}  &  {\tiny{}$0.5$}  &  {\tiny{}$0.066$}  &  {\tiny{}$1.0$}  &  {\tiny{}$0.064$}  &  {\tiny{}$2.0$}  &  {\tiny{}$0.047$}  &  {\tiny{}$2.028$}  &  {\tiny{}$0.044$}   & \ldots & \ldots  \tabularnewline
  \ldots  & \ldots & \ldots & \ldots  &  {\tiny{}$10.0$}  &  {\tiny{}$0.2$}  &  {\tiny{}$>0.35$}  &  {\tiny{}$0.5$}  &  {\tiny{}$0.349$}  &  {\tiny{}$1.0$}  &  {\tiny{}$0.121$}  &  {\tiny{}$2.0$}  &  {\tiny{}$0.053$}  &  {\tiny{}$2.028$}  &  {\tiny{}$0.049$}  & \ldots & \ldots  \tabularnewline
{\tiny{}$0.725$}  &  {\tiny{}$39.3$}  &  {\tiny{}$0.103$}  &  {\tiny{}$0.086$}  &  {\tiny{}$0.01$}  &  {\tiny{}$0.2$}  &  {\tiny{}$>0.35$}  &  {\tiny{}$0.5$}  &  {\tiny{}$0.231$}  &  {\tiny{}$1.0$}  &  {\tiny{}$0.108$}  &  {\tiny{}$2.0$}  &  {\tiny{}$0.052$}  &  {\tiny{}$2.102$}  &  {\tiny{}$0.046$}   & \ldots & \ldots  \tabularnewline
  \ldots  & \ldots & \ldots & \ldots  &  {\tiny{}$0.1$}  &  {\tiny{}$0.2$}  &  {\tiny{}$>0.35$}  &  {\tiny{}$0.5$}  &  {\tiny{}$0.148$}  &  {\tiny{}$1.0$}  &  {\tiny{}$0.060$}  &  {\tiny{}$2.0$}  &  {\tiny{}$0.031$}  &  {\tiny{}$2.102$}  &  {\tiny{}$0.027$}   & \ldots & \ldots  \tabularnewline
  \ldots  & \ldots & \ldots & \ldots  &  {\tiny{}$1.0$}  &  {\tiny{}$0.2$}  &  {\tiny{}$0.176$}  &  {\tiny{}$0.5$}  &  {\tiny{}$0.064$}  &  {\tiny{}$1.0$}  &  {\tiny{}$0.038$}  &  {\tiny{}$2.0$}  &  {\tiny{}$0.036$}  &  {\tiny{}$2.102$}  &  {\tiny{}$0.035$}   & \ldots & \ldots  \tabularnewline
  \ldots  & \ldots & \ldots & \ldots  &  {\tiny{}$ 8.0$}  &  {\tiny{}$0.2$}  &  {\tiny{}$0.228$}  &  {\tiny{}$0.5$}  &  {\tiny{}$0.210$}  &  {\tiny{}$1.0$}  &  {\tiny{}$0.107$}  &  {\tiny{}$2.0$}  &  {\tiny{}$0.051$}  &  {\tiny{}$2.102$}  &  {\tiny{}$0.046$}  & \ldots & \ldots  \tabularnewline
{\tiny{}$0.750$}  &  {\tiny{}$35.2$}  &  {\tiny{}$0.098$}  &  {\tiny{}$0.083$}  &  {\tiny{}$0.01$}  &  {\tiny{}$0.2$}  &  {\tiny{}$>0.35$}  &  {\tiny{}$0.5$}  &  {\tiny{}$0.233$}  &  {\tiny{}$1.0$}  &  {\tiny{}$0.108$}  &  {\tiny{}$2.0$}  &  {\tiny{}$0.052$}  &  {\tiny{}$2.178$}  &  {\tiny{}$0.045$}   & \ldots & \ldots  \tabularnewline
  \ldots  & \ldots & \ldots & \ldots  &  {\tiny{}$0.1$}  &  {\tiny{}$0.2$}  &  {\tiny{}$>0.35$}  &  {\tiny{}$0.5$}  &  {\tiny{}$0.168$}  &  {\tiny{}$1.0$}  &  {\tiny{}$0.064$}  &  {\tiny{}$2.0$}  &  {\tiny{}$0.032$}  &  {\tiny{}$2.178$}  &  {\tiny{}$0.026$}   & \ldots & \ldots  \tabularnewline
  \ldots  & \ldots & \ldots & \ldots  &  {\tiny{}$1.0$}  &  {\tiny{}$0.2$}  &  {\tiny{}$0.215$}  &  {\tiny{}$0.5$}  &  {\tiny{}$0.069$}  &  {\tiny{}$1.0$}  &  {\tiny{}$0.037$}  &  {\tiny{}$2.0$}  &  {\tiny{}$0.022$}  &  {\tiny{}$2.178$}  &  {\tiny{}$0.020$}   & \ldots & \ldots  \tabularnewline
  \ldots  & \ldots & \ldots & \ldots  &  {\tiny{}$10.0$}  &  {\tiny{}$0.2$}  &  {\tiny{}$0.142$}  &  {\tiny{}$0.5$}  &  {\tiny{}$0.143$}  &  {\tiny{}$1.0$}  &  {\tiny{}$0.093$}  &  {\tiny{}$2.0$}  &  {\tiny{}$0.048$}  &  {\tiny{}$2.178$}  &  {\tiny{}$0.042$}  & \ldots & \ldots  \tabularnewline
{\tiny{}$0.775$}  &  {\tiny{}$31.6$}  &  {\tiny{}$0.092$}  &  {\tiny{}$0.080$}  &  {\tiny{}$0.01$}  &  {\tiny{}$0.2$}  &  {\tiny{}$>0.35$}  &  {\tiny{}$0.5$}  &  {\tiny{}$0.237$}  &  {\tiny{}$1.0$}  &  {\tiny{}$0.107$}  &  {\tiny{}$2.0$}  &  {\tiny{}$0.052$}  &  {\tiny{}$2.255$}  &  {\tiny{}$0.044$}   & \ldots & \ldots  \tabularnewline
  \ldots  & \ldots & \ldots & \ldots  &  {\tiny{}$0.1$}  &  {\tiny{}$0.2$}  &  {\tiny{}$>0.35$}  &  {\tiny{}$0.5$}  &  {\tiny{}$0.192$}  &  {\tiny{}$1.0$}  &  {\tiny{}$0.070$}  &  {\tiny{}$2.0$}  &  {\tiny{}$0.034$}  &  {\tiny{}$2.255$}  &  {\tiny{}$0.026$}   & \ldots & \ldots  \tabularnewline
  \ldots  & \ldots & \ldots & \ldots  &  {\tiny{}$1.0$}  &  {\tiny{}$0.2$}  &  {\tiny{}$0.278$}  &  {\tiny{}$0.5$}  &  {\tiny{}$0.077$}  &  {\tiny{}$1.0$}  &  {\tiny{}$0.040$}  &  {\tiny{}$2.0$}  &  {\tiny{}$0.022$}  &  {\tiny{}$2.255$}  &  {\tiny{}$0.018$}   & \ldots & \ldots  \tabularnewline
  \ldots  & \ldots & \ldots & \ldots  &  {\tiny{}$ 8.0$}  &  {\tiny{}$0.2$}  &  {\tiny{}$0.067$}  &  {\tiny{}$0.5$}  &  {\tiny{}$0.061$}  &  {\tiny{}$1.0$}  &  {\tiny{}$0.061$}  &  {\tiny{}$2.0$}  &  {\tiny{}$0.042$}  &  {\tiny{}$2.255$}  &  {\tiny{}$0.036$}  & \ldots & \ldots  \tabularnewline
{\tiny{}$0.800$}  &  {\tiny{}$28.4$}  &  {\tiny{}$0.087$}  &  {\tiny{}$0.077$}  &  {\tiny{}$0.01$}  &  {\tiny{}$0.2$}  &  {\tiny{}$>0.35$}  &  {\tiny{}$0.5$}  &  {\tiny{}$0.251$}  &  {\tiny{}$1.0$}  &  {\tiny{}$0.107$}  &  {\tiny{}$2.0$}  &  {\tiny{}$0.052$}  &  {\tiny{}$2.334$}  &  {\tiny{}$0.042$}   & \ldots & \ldots  \tabularnewline
  \ldots  & \ldots & \ldots & \ldots  &  {\tiny{}$0.1$}  &  {\tiny{}$0.2$}  &  {\tiny{}$>0.35$}  &  {\tiny{}$0.5$}  &  {\tiny{}$0.220$}  &  {\tiny{}$1.0$}  &  {\tiny{}$0.076$}  &  {\tiny{}$2.0$}  &  {\tiny{}$0.036$}  &  {\tiny{}$2.334$}  &  {\tiny{}$0.027$}   & \ldots & \ldots  \tabularnewline
  \ldots  & \ldots & \ldots & \ldots  &  {\tiny{}$1.0$}  &  {\tiny{}$0.2$}  &  {\tiny{}$>0.35$}  &  {\tiny{}$0.5$}  &  {\tiny{}$0.085$}  &  {\tiny{}$1.0$}  &  {\tiny{}$0.044$}  &  {\tiny{}$2.0$}  &  {\tiny{}$0.024$}  &  {\tiny{}$2.334$}  &  {\tiny{}$0.018$}   & \ldots & \ldots  \tabularnewline
  \ldots  & \ldots & \ldots & \ldots  &  {\tiny{}$10.0$}  &  {\tiny{}$0.2$}  &  {\tiny{}$  -  $}  &  {\tiny{}$0.5$}  &  {\tiny{}$  -  $}  &  {\tiny{}$1.0$}  &  {\tiny{}$  -  $}  &  {\tiny{}$2.0$}  &  {\tiny{}$0.024$}  &  {\tiny{}$2.334$}  &  {\tiny{}$0.024$}  & \ldots & \ldots  \tabularnewline
{\tiny{}$0.825$}  &  {\tiny{}$25.6$}  &  {\tiny{}$0.082$}  &  {\tiny{}$0.075$}  &  {\tiny{}$0.01$}  &  {\tiny{}$0.2$}  &  {\tiny{}$>0.35$}  &  {\tiny{}$0.5$}  &  {\tiny{}$0.271$}  &  {\tiny{}$1.0$}  &  {\tiny{}$0.108$}  &  {\tiny{}$2.0$}  &  {\tiny{}$0.052$}  &  {\tiny{}$2.412$}  &  {\tiny{}$0.041$}   & \ldots & \ldots  \tabularnewline
  \ldots  & \ldots & \ldots & \ldots  &  {\tiny{}$0.1$}  &  {\tiny{}$0.2$}  &  {\tiny{}$>0.35$}  &  {\tiny{}$0.5$}  &  {\tiny{}$0.247$}  &  {\tiny{}$1.0$}  &  {\tiny{}$0.082$}  &  {\tiny{}$2.0$}  &  {\tiny{}$0.038$}  &  {\tiny{}$2.412$}  &  {\tiny{}$0.027$}   & \ldots & \ldots  \tabularnewline
  \ldots  & \ldots & \ldots & \ldots  &  {\tiny{}$1.0$}  &  {\tiny{}$0.2$}  &  {\tiny{}$>0.35$}  &  {\tiny{}$0.5$}  &  {\tiny{}$0.093$}  &  {\tiny{}$1.0$}  &  {\tiny{}$0.048$}  &  {\tiny{}$2.0$}  &  {\tiny{}$0.026$}  &  {\tiny{}$2.412$}  &  {\tiny{}$0.019$}   & \ldots & \ldots  \tabularnewline
  \ldots  & \ldots & \ldots & \ldots  &  {\tiny{}$10.0$}  &  {\tiny{}$0.2$}  &  {\tiny{}$0.059$}  &  {\tiny{}$0.5$}  &  {\tiny{}$0.033$}  &  {\tiny{}$1.0$}  &  {\tiny{}$0.021$}  &  {\tiny{}$2.0$}  &  {\tiny{}$0.014$}  &  {\tiny{}$2.412$}  &  {\tiny{}$0.011$}  & \ldots & \ldots  \tabularnewline
\multicolumn{17}{c}{{\tiny{}$Z=0.008$; $M_\text{AGB}=3.0\ M_{\odot}$; $t_\mathrm{mt}=422\ \mathrm{Myr}$; uniform rotation}}\tabularnewline
{\tiny{}$1.000$}  &  {\tiny{}$25.27$} &  {\tiny{}$0.075$}  &  {\tiny{}$0.082$}  &   {\tiny{}$0.01$}  &  {\tiny{}$0.2$}  &   {\tiny{}$>2.0$}  &  {\tiny{}$0.5$}  &  {\tiny{}$0.275$}  &  {\tiny{}$1.0$}  &  {\tiny{}$0.136$}  &  {\tiny{}$2.0$}  &  {\tiny{}$0.070$}  &   {\tiny{}$3.0$}   &  {\tiny{}$0.045$}  &  {\tiny{}$2.954$}  &  {\tiny{}$0.045$}  \tabularnewline
  \ldots  & \ldots & \ldots & \ldots  &  {\tiny{}$0.1$}  &  {\tiny{}$0.2$}  &   {\tiny{}$>2.0$}  &  {\tiny{}$0.5$}  &  {\tiny{}$0.196$}  &  {\tiny{}$1.0$}  &  {\tiny{}$0.088$}  &  {\tiny{}$2.0$}  &  {\tiny{}$0.047$}  &   {\tiny{}$3.0$}   &  {\tiny{}$0.033$}  &  {\tiny{}$2.954$}  &  {\tiny{}$0.031$}  \tabularnewline
  \ldots  & \ldots & \ldots & \ldots  &  {\tiny{}$1.0$}  &  {\tiny{}$0.2$}  &  {\tiny{}$0.255$}  &  {\tiny{}$0.5$}  &  {\tiny{}$0.086$}  &  {\tiny{}$1.0$}  &  {\tiny{}$0.046$}  &  {\tiny{}$2.0$}  &  {\tiny{}$0.029$}  &  {\tiny{}$3.0$}  &  {\tiny{}$0.029$}  &  {\tiny{}$2.954$}  &  {\tiny{}$0.029$}  \tabularnewline
  \ldots  & \ldots & \ldots & \ldots  &  {\tiny{}$10.0$}  &  {\tiny{}$0.2$}  &  {\tiny{}$0.181$}  &  {\tiny{}$0.5$}  &  {\tiny{}$0.181$}  &  {\tiny{}$1.0$}  &  {\tiny{}$0.127$}  &  {\tiny{}$2.0$}  &  {\tiny{}$0.067$}  &  {\tiny{}$3.0$}  &  {\tiny{}$0.045$}  &  {\tiny{}$2.954$}  &  {\tiny{}$0.043$}  \tabularnewline
{\tiny{}$1.250$}  &  {\tiny{}$12.60$} &  {\tiny{}$0.045$}  &  {\tiny{}$0.059$}  &   {\tiny{}$0.01$}  &  {\tiny{}$0.2$}  &   {\tiny{}$>2.0$}  &  {\tiny{}$0.5$}  &  {\tiny{}$0.490$}  &  {\tiny{}$1.0$}  &  {\tiny{}$0.160$}  &  {\tiny{}$2.0$}  &  {\tiny{}$0.071$}  &   {\tiny{}$3.0$}   &  {\tiny{}$0.046$}  &  {\tiny{}$3.844$}  &  {\tiny{}$0.035$}  \tabularnewline
  \ldots  & \ldots & \ldots & \ldots  &  {\tiny{}$0.1$}  &  {\tiny{}$0.2$}  &   {\tiny{}$>2.0$}  &  {\tiny{}$0.5$}  &  {\tiny{}$0.410$}  &  {\tiny{}$1.0$}  &  {\tiny{}$0.139$}  &  {\tiny{}$2.0$}  &  {\tiny{}$0.062$}  &   {\tiny{}$3.0$}   &  {\tiny{}$0.041$}  &  {\tiny{}$3.844$}  &  {\tiny{}$0.030$}  \tabularnewline
  \ldots  & \ldots & \ldots & \ldots  &  {\tiny{}$1.0$}  &  {\tiny{}$0.2$}  &  {\tiny{}$>2.0 $}  &  {\tiny{}$0.5$}  &  {\tiny{}$0.174$}  &  {\tiny{}$1.0$}  &  {\tiny{}$0.083$}  &  {\tiny{}$2.0$}  &  {\tiny{}$0.044$}  &  {\tiny{}$3.0$}  &  {\tiny{}$0.031$}  &  {\tiny{}$3.844$}  &  {\tiny{}$0.022$}  \tabularnewline
  \ldots  & \ldots & \ldots & \ldots  &  {\tiny{}$10.0$}  &  {\tiny{}$0.2$}  &  {\tiny{}$  -  $}  &  {\tiny{}$0.5$}  &  {\tiny{}$  -  $}  &  {\tiny{}$1.0$}  &  {\tiny{}$  -  $}  &  {\tiny{}$2.0$}  &  {\tiny{}$  -  $}  &  {\tiny{}$3.0$}  &  {\tiny{}$  -  $}  &  {\tiny{}$3.844$}  &  {\tiny{}$  -  $}  \tabularnewline
{\tiny{}$1.500$}  &  {\tiny{}$ 7.27$} &  {\tiny{}$0.044$}  &  {\tiny{}$0.069$}  &   {\tiny{}$0.01$}  &  {\tiny{}$0.2$}  &   {\tiny{}$>2.0$}  &  {\tiny{}$0.5$}  &  {\tiny{}$1.052$}  &  {\tiny{}$1.0$}  &  {\tiny{}$0.240$}  &  {\tiny{}$2.0$}  &  {\tiny{}$0.102$}  &   {\tiny{}$3.0$}   &  {\tiny{}$0.065$}  &  {\tiny{}$4.370$}  &  {\tiny{}$0.042$}  \tabularnewline
  \ldots  & \ldots & \ldots & \ldots  &  {\tiny{}$0.1$}  &  {\tiny{}$0.2$}  &   {\tiny{}$>2.0$}  &  {\tiny{}$0.5$}  &  {\tiny{}$1.041$}  &  {\tiny{}$1.0$}  &  {\tiny{}$0.225$}  &  {\tiny{}$2.0$}  &  {\tiny{}$0.094$}  &   {\tiny{}$3.0$}   &  {\tiny{}$0.060$}  &  {\tiny{}$4.370$}  &  {\tiny{}$0.039$}  \tabularnewline
  \ldots  & \ldots & \ldots & \ldots  &  {\tiny{}$1.0$}  &  {\tiny{}$0.2$}  &  {\tiny{}$>2.0 $}  &  {\tiny{}$0.5$}  &  {\tiny{}$0.365$}  &  {\tiny{}$1.0$}  &  {\tiny{}$0.139$}  &  {\tiny{}$2.0$}  &  {\tiny{}$0.069$}  &  {\tiny{}$3.0$}  &  {\tiny{}$0.047$}  &  {\tiny{}$4.370$}  &  {\tiny{}$0.030$}  \tabularnewline
  \ldots  & \ldots & \ldots & \ldots  &  {\tiny{}$10.0$}  &  {\tiny{}$0.2$}  &  {\tiny{}$0.286$}  &  {\tiny{}$0.5$}  &  {\tiny{}$  -  $}  &  {\tiny{}$1.0$}  &  {\tiny{}$0.069$}  &  {\tiny{}$2.0$}  &  {\tiny{}$  -  $}  &  {\tiny{}$3.0$}  &  {\tiny{}$  -  $}  &  {\tiny{}$4.370$}  &  {\tiny{}$  -  $}  \tabularnewline
{\tiny{}$1.735$}  &  {\tiny{}$ 4.63$} &  {\tiny{}$0.044$}  &  {\tiny{}$0.080$}  &   {\tiny{}$0.01$}  &  {\tiny{}$0.2$}  &   {\tiny{}$>2.0$}  &  {\tiny{}$0.5$}  &  {\tiny{}$>2.0 $}  &  {\tiny{}$1.0$}  &  {\tiny{}$0.338$}  &  {\tiny{}$2.0$}  &  {\tiny{}$0.137$}  &   {\tiny{}$3.0$}   &  {\tiny{}$0.087$}  &  {\tiny{}$4.954$}  &  {\tiny{}$0.049$}  \tabularnewline
  \ldots  & \ldots & \ldots & \ldots  &  {\tiny{}$0.1$}  &  {\tiny{}$0.2$}  &   {\tiny{}$>2.0$}  &  {\tiny{}$0.5$}  &  {\tiny{}$>2.0 $}  &  {\tiny{}$1.0$}  &  {\tiny{}$0.327$}  &  {\tiny{}$2.0$}  &  {\tiny{}$0.133$}  &   {\tiny{}$3.0$}   &  {\tiny{}$0.083$}  &  {\tiny{}$4.954$}  &  {\tiny{}$0.047$}  \tabularnewline
  \ldots  & \ldots & \ldots & \ldots  &  {\tiny{}$1.0$}  &  {\tiny{}$0.2$}  &  {\tiny{}$>2.0 $}  &  {\tiny{}$0.5$}  &  {\tiny{}$>2.0 $}  &  {\tiny{}$1.0$}  &  {\tiny{}$0.218$}  &  {\tiny{}$2.0$}  &  {\tiny{}$0.100$}  &  {\tiny{}$3.0$}  &  {\tiny{}$0.066$}  &  {\tiny{}$4.954$}  &  {\tiny{}$0.038$}  \tabularnewline
  \ldots  & \ldots & \ldots & \ldots  &  {\tiny{}$ 8.0$}  &  {\tiny{}$0.2$}  &  {\tiny{}$>2.0 $}  &  {\tiny{}$0.5$}  &  {\tiny{}$0.234$}  &  {\tiny{}$1.0$}  &  {\tiny{}$0.117$}  &  {\tiny{}$2.0$}  &  {\tiny{}$0.064$}  &  {\tiny{}$3.0$}  &  {\tiny{}$0.045$}  &  {\tiny{}$4.954$}  &  {\tiny{}$0.027$}  \tabularnewline
{\tiny{}$2.000$}  &  {\tiny{}$ 2.84$} &  {\tiny{}$0.042$}  &  {\tiny{}$0.089$}  &   {\tiny{}$0.1$}  &  {\tiny{}$0.2$}  &   {\tiny{}$>2.0$}  &  {\tiny{}$0.5$}  &  {\tiny{}$>2.0 $}  &  {\tiny{}$1.0$}  &  {\tiny{}$0.470$}  &  {\tiny{}$2.0$}  &  {\tiny{}$0.180$}  &   {\tiny{}$3.0$}   &  {\tiny{}$0.112$}  &  {\tiny{}$5.770$}  &  {\tiny{}$0.054$}  \tabularnewline
  \ldots  & \ldots & \ldots & \ldots  &  {\tiny{}$1.0$}  &  {\tiny{}$0.2$}  &  {\tiny{}$>2.0 $}  &  {\tiny{}$0.5$}  &  {\tiny{}$>2.0 $}  &  {\tiny{}$1.0$}  &  {\tiny{}$0.358$}  &  {\tiny{}$2.0$}  &  {\tiny{}$0.143$}  &  {\tiny{}$3.0$}  &  {\tiny{}$0.092$}  &  {\tiny{}$5.770$}  &  {\tiny{}$0.045$}  \tabularnewline
  \ldots  & \ldots & \ldots & \ldots  &  {\tiny{}$10.0$}  &  {\tiny{}$0.2$}  &  {\tiny{}$>2.0 $}  &  {\tiny{}$0.5$}  &  {\tiny{}$0.379$}  &  {\tiny{}$1.0$}  &  {\tiny{}$0.171$}  &  {\tiny{}$2.0$}  &  {\tiny{}$0.089$}  &  {\tiny{}$3.0$}  &  {\tiny{}$0.062$}  &  {\tiny{}$5.770$}  &  {\tiny{}$0.032$}  \tabularnewline
{\tiny{}$2.250$}  &  {\tiny{}$ 1.77$} &  {\tiny{}$0.039$}  &  {\tiny{}$0.093$}  &   {\tiny{}$0.1$}  &  {\tiny{}$0.2$}  &   {\tiny{}$>2.0$}  &  {\tiny{}$0.5$}  &  {\tiny{}$>2.0 $}  &  {\tiny{}$1.0$}  &  {\tiny{}$0.635$}  &  {\tiny{}$2.0$}  &  {\tiny{}$0.229$}  &   {\tiny{}$3.0$}   &  {\tiny{}$0.141$}  &  {\tiny{}$6.809$}  &  {\tiny{}$0.056$}  \tabularnewline
  \ldots  & \ldots & \ldots & \ldots  &  {\tiny{}$1.0$}  &  {\tiny{}$0.2$}  &  {\tiny{}$>2.0 $}  &  {\tiny{}$0.5$}  &  {\tiny{}$>2.0 $}  &  {\tiny{}$1.0$}  &  {\tiny{}$0.535$}  &  {\tiny{}$2.0$}  &  {\tiny{}$0.191$}  &  {\tiny{}$3.0$}  &  {\tiny{}$0.120$}  &  {\tiny{}$6.809$}  &  {\tiny{}$0.049$}  \tabularnewline
  \ldots  & \ldots & \ldots & \ldots  &  {\tiny{}$10.0$}  &  {\tiny{}$0.2$}  &  {\tiny{}$>2.0 $}  &  {\tiny{}$0.5$}  &  {\tiny{}$0.732$}  &  {\tiny{}$1.0$}  &  {\tiny{}$0.253$}  &  {\tiny{}$2.0$}  &  {\tiny{}$0.124$}  &  {\tiny{}$3.0$}  &  {\tiny{}$0.085$}  &  {\tiny{}$6.809$}  &  {\tiny{}$0.037$}  \tabularnewline
{\tiny{}$2.500$}  &  {\tiny{}$ 1.07$} &  {\tiny{}$0.034$}  &  {\tiny{}$0.088$}  &   {\tiny{}$0.1$}  &  {\tiny{}$0.2$}  &   {\tiny{}$>2.0$}  &  {\tiny{}$0.5$}  &  {\tiny{}$>2.0 $}  &  {\tiny{}$1.0$}  &  {\tiny{}$0.746$}  &  {\tiny{}$2.0$}  &  {\tiny{}$0.278$}  &   {\tiny{}$3.0$}   &  {\tiny{}$0.170$}  &  {\tiny{}$8.460$}  &  {\tiny{}$0.054$}  \tabularnewline
  \ldots  & \ldots & \ldots & \ldots  &  {\tiny{}$1.0$}  &  {\tiny{}$0.2$}  &  {\tiny{}$>2.0 $}  &  {\tiny{}$0.5$}  &  {\tiny{}$>2.0 $}  &  {\tiny{}$1.0$}  &  {\tiny{}$0.728$}  &  {\tiny{}$2.0$}  &  {\tiny{}$0.246$}  &  {\tiny{}$3.0$}  &  {\tiny{}$0.149$}  &  {\tiny{}$8.460$}  &  {\tiny{}$0.048$}  \tabularnewline
  \ldots  & \ldots & \ldots & \ldots  &  {\tiny{}$10.0$}  &  {\tiny{}$0.2$}  &  {\tiny{}$>2.0 $}  &  {\tiny{}$0.5$}  &  {\tiny{}$>2.0 $}  &  {\tiny{}$1.0$}  &  {\tiny{}$0.373$}  &  {\tiny{}$2.0$}  &  {\tiny{}$0.168$}  &  {\tiny{}$3.0$}  &  {\tiny{}$0.112$}  &  {\tiny{}$8.460$}  &  {\tiny{}$0.039$}  \tabularnewline
\multicolumn{17}{c}{{\tiny{}$Z=10^{-4}$; $M_\text{AGB}=1.0\ M_{\odot}$; $t_\mathrm{mt}=6.3\ \mathrm{Gyr}$; differential rotation}}\tabularnewline
{\tiny{}$0.600$}  &  {\tiny{}$65.0$}  &  {\tiny{}$0.128$}  &  {\tiny{}$0.091$}  &  {\tiny{}$0.01$}  &  {\tiny{}$0.2$}  &  {\tiny{}$>0.35$}  &  {\tiny{}$0.5$}  &  {\tiny{}$>0.35$}  &  {\tiny{}$1.0$}  &  {\tiny{}$  -  $}  &  {\tiny{}$2.0$}  &  {\tiny{}$  -  $}  &  {\tiny{}$1.758$}  &  {\tiny{}$  -  $}   & \ldots & \ldots  \tabularnewline
  \ldots  & \ldots & \ldots & \ldots  &  {\tiny{}$0.1$}  &  {\tiny{}$0.2$}  &  {\tiny{}$>0.35$}  &  {\tiny{}$0.5$}  &  {\tiny{}$>0.35$}  &  {\tiny{}$1.0$}  &  {\tiny{}$0.004$}  &  {\tiny{}$2.0$}  &  {\tiny{}$0.004$}  &  {\tiny{}$1.758$}  &  {\tiny{}$0.004$}   & \ldots & \ldots  \tabularnewline
  \ldots  & \ldots & \ldots & \ldots  &  {\tiny{}$1.0$}  &  {\tiny{}$0.2$}  &  {\tiny{}$  -  $}  &  {\tiny{}$0.5$}  &  {\tiny{}$0.160$}  &  {\tiny{}$0.7$}  &  {\tiny{}$0.072$}  &  {\tiny{}$1.0$}  &  {\tiny{}$0.001$}  &    {\tiny{}$2.0$}  &  {\tiny{}$  -  $}  &  {\tiny{}$1.758$}  &  {\tiny{}$0.001$}  \tabularnewline
  \ldots  & \ldots & \ldots & \ldots  &  {\tiny{}$10.0$}  &  {\tiny{}$0.2$}  &  {\tiny{}$  -  $}  &  {\tiny{}$0.5$}  &  {\tiny{}$  -  $}  &  {\tiny{}$1.0$}  &  {\tiny{}$  -  $}  &  {\tiny{}$2.0$}  &  {\tiny{}$  -  $}  &  {\tiny{}$1.758$}  &  {\tiny{}$  -  $}  & \ldots & \ldots  \tabularnewline
{\tiny{}$0.625$}  &  {\tiny{}$56.2$}  &  {\tiny{}$0.120$}  &  {\tiny{}$0.088$}  &  {\tiny{}$0.01$}  &  {\tiny{}$0.2$}  &  {\tiny{}$>0.35$}  &  {\tiny{}$0.5$}  &  {\tiny{}$>0.35$}  &  {\tiny{}$1.0$}  &  {\tiny{}$0.000$}  &  {\tiny{}$2.0$}  &  {\tiny{}$  -  $}  &  {\tiny{}$1.836$}  &  {\tiny{}$  -  $}   & \ldots & \ldots  \tabularnewline
  \ldots  & \ldots & \ldots & \ldots  &  {\tiny{}$0.1$}  &  {\tiny{}$0.2$}  &  {\tiny{}$>0.35$}  &  {\tiny{}$0.5$}  &  {\tiny{}$>0.35$}  &  {\tiny{}$1.0$}  &  {\tiny{}$0.000$}  &  {\tiny{}$2.0$}  &  {\tiny{}$0.000$}  &  {\tiny{}$1.836$}  &  {\tiny{}$0.000$}   & \ldots & \ldots  \tabularnewline
  \ldots  & \ldots & \ldots & \ldots  &  {\tiny{}$1.0$}  &  {\tiny{}$0.2$}  &  {\tiny{}$>0.35$}  &  {\tiny{}$0.5$}  &  {\tiny{}$>0.35$}  &  {\tiny{}$0.7$}  &  {\tiny{}$0.056$}  &  {\tiny{}$1.0$}  &  {\tiny{}$0.005$}  &    {\tiny{}$2.0$}  &  {\tiny{}$0.005$}  &  {\tiny{}$1.836$}  &  {\tiny{}$0.005$}  \tabularnewline
  \ldots  & \ldots & \ldots & \ldots  &  {\tiny{}$10.0$}  &  {\tiny{}$0.2$}  &  {\tiny{}$  -  $}  &  {\tiny{}$0.5$}  &  {\tiny{}$  -  $}  &  {\tiny{}$1.0$}  &  {\tiny{}$  -  $}  &  {\tiny{}$2.0$}  &  {\tiny{}$  -  $}  &  {\tiny{}$1.836$}  &  {\tiny{}$  -  $}  & \ldots & \ldots  \tabularnewline
{\tiny{}$0.650$}  &  {\tiny{}$48.7$}  &  {\tiny{}$0.112$}  &  {\tiny{}$0.085$}  &  {\tiny{}$0.01$}  &  {\tiny{}$0.2$}  &  {\tiny{}$>0.35$}  &  {\tiny{}$0.5$}  &  {\tiny{}$>0.35$}  &  {\tiny{}$1.0$}  &  {\tiny{}$0.000$}  &  {\tiny{}$2.0$}  &  {\tiny{}$0.000$}  &  {\tiny{}$1.914$}  &  {\tiny{}$0.000$}   & \ldots & \ldots  \tabularnewline
  \ldots  & \ldots & \ldots & \ldots  &  {\tiny{}$0.1$}  &  {\tiny{}$0.2$}  &  {\tiny{}$>0.35$}  &  {\tiny{}$0.5$}  &  {\tiny{}$>0.35$}  &  {\tiny{}$1.0$}  &  {\tiny{}$0.000$}  &  {\tiny{}$2.0$}  &  {\tiny{}$0.000$}  &  {\tiny{}$1.914$}  &  {\tiny{}$0.000$}   & \ldots & \ldots  \tabularnewline
  \ldots  & \ldots & \ldots & \ldots  &  {\tiny{}$1.0$}  &  {\tiny{}$0.2$}  &  {\tiny{}$>0.35$}  &  {\tiny{}$0.5$}  &  {\tiny{}$>0.35$}  &  {\tiny{}$0.7$}  &  {\tiny{}$0.050$}  &  {\tiny{}$1.0$}  &  {\tiny{}$0.006$}  &    {\tiny{}$2.0$}  &  {\tiny{}$0.001$}  &  {\tiny{}$1.914$}  &  {\tiny{}$0.001$}  \tabularnewline
  \ldots  & \ldots & \ldots & \ldots  &  {\tiny{}$10.0$}  &  {\tiny{}$0.2$}  &  {\tiny{}$  -  $}  &  {\tiny{}$0.5$}  &  {\tiny{}$  -  $}  &  {\tiny{}$1.0$}  &  {\tiny{}$  -  $}  &  {\tiny{}$2.0$}  &  {\tiny{}$  -  $}  &  {\tiny{}$1.914$}  &  {\tiny{}$  -  $}  & \ldots & \ldots \tabularnewline
{\tiny{}$0.675$}  &  {\tiny{}$42.2$}  &  {\tiny{}$0.105$}  &  {\tiny{}$0.081$}  &  {\tiny{}$0.01$}  &  {\tiny{}$0.2$}  &  {\tiny{}$>0.35$}  &  {\tiny{}$0.5$}  &  {\tiny{}$>0.35$}  &  {\tiny{}$1.0$}  &  {\tiny{}$0.000$}  &  {\tiny{}$2.0$}  &  {\tiny{}$  -  $}  &  {\tiny{}$1.995$}  &  {\tiny{}$  -  $}   & \ldots & \ldots  \tabularnewline
  \ldots  & \ldots & \ldots & \ldots  &  {\tiny{}$0.1$}  &  {\tiny{}$0.2$}  &  {\tiny{}$>0.35$}  &  {\tiny{}$0.5$}  &  {\tiny{}$>0.35$}  &  {\tiny{}$1.0$}  &  {\tiny{}$0.000$}  &  {\tiny{}$2.0$}  &  {\tiny{}$0.000$}  &  {\tiny{}$1.995$}  &  {\tiny{}$0.000$}   & \ldots & \ldots  \tabularnewline
  \ldots  & \ldots & \ldots & \ldots  &  {\tiny{}$1.0$}  &  {\tiny{}$0.2$}  &  {\tiny{}$>0.35$}  &  {\tiny{}$0.5$}  &  {\tiny{}$>0.35$}  &  {\tiny{}$0.7$}  &  {\tiny{}$0.039$}  &  {\tiny{}$1.0$}  &  {\tiny{}$0.000$}  &    {\tiny{}$2.0$}  &  {\tiny{}$0.000$}  &  {\tiny{}$1.995$}  &  {\tiny{}$0.000$}  \tabularnewline
  \ldots  & \ldots & \ldots & \ldots  &  {\tiny{}$10.0$}  &  {\tiny{}$0.2$}  &  {\tiny{}$  -  $}  &  {\tiny{}$0.5$}  &  {\tiny{}$  -  $}  &  {\tiny{}$1.0$}  &  {\tiny{}$  -  $}  &  {\tiny{}$2.0$}  &  {\tiny{}$  -  $}  &  {\tiny{}$1.995$}  &  {\tiny{}$  -  $}  & \ldots & \ldots  \tabularnewline
{\tiny{}$0.700$}  &  {\tiny{}$36.5$}  &  {\tiny{}$0.097$}  &  {\tiny{}$0.077$}  &  {\tiny{}$0.01$}  &  {\tiny{}$0.2$}  &  {\tiny{}$>0.35$}  &  {\tiny{}$0.5$}  &  {\tiny{}$>0.35$}  &  {\tiny{}$1.0$}  &  {\tiny{}$0.000$}  &  {\tiny{}$2.0$}  &  {\tiny{}$  -  $}  &  {\tiny{}$2.080$}  &  {\tiny{}$  -  $}   & \ldots & \ldots  \tabularnewline
  \ldots  & \ldots & \ldots & \ldots  &  {\tiny{}$0.1$}  &  {\tiny{}$0.2$}  &  {\tiny{}$>0.35$}  &  {\tiny{}$0.5$}  &  {\tiny{}$  -  $}  &  {\tiny{}$1.0$}  &  {\tiny{}$0.000$}  &  {\tiny{}$2.0$}  &  {\tiny{}$0.000$}  &  {\tiny{}$2.080$}  &  {\tiny{}$0.000$}   & \ldots & \ldots  \tabularnewline
  \ldots  & \ldots & \ldots & \ldots  &  {\tiny{}$1.0$}  &  {\tiny{}$0.2$}  &  {\tiny{}$>0.35$}  &  {\tiny{}$0.5$}  &  {\tiny{}$>0.35$}  &  {\tiny{}$0.7$}  &  {\tiny{}$0.029$}  &  {\tiny{}$1.0$}  &  {\tiny{}$0.000$}  &    {\tiny{}$2.0$}  &  {\tiny{}$0.000$}  &  {\tiny{}$2.080$}  &  {\tiny{}$0.000$}  \tabularnewline
  \ldots  & \ldots & \ldots & \ldots  &  {\tiny{}$10.0$}  &  {\tiny{}$0.2$}  &  {\tiny{}$  -  $}  &  {\tiny{}$0.5$}  &  {\tiny{}$  -  $}  &  {\tiny{}$1.0$}  &  {\tiny{}$  -  $}  &  {\tiny{}$2.0$}  &  {\tiny{}$  -  $}  &  {\tiny{}$2.080$}  &  {\tiny{}$  -  $}  & \ldots & \ldots  \tabularnewline
{\tiny{}$0.725$}  &  {\tiny{}$31.6$}  &  {\tiny{}$0.089$}  &  {\tiny{}$0.073$}  &  {\tiny{}$0.01$}  &  {\tiny{}$0.2$}  &  {\tiny{}$>0.35$}  &  {\tiny{}$0.5$}  &  {\tiny{}$>0.35$}  &  {\tiny{}$1.0$}  &  {\tiny{}$0.000$}  &  {\tiny{}$2.0$}  &  {\tiny{}$  -  $}  &  {\tiny{}$2.168$}  &  {\tiny{}$  -  $}   & \ldots & \ldots  \tabularnewline
  \ldots  & \ldots & \ldots & \ldots  &  {\tiny{}$0.1$}  &  {\tiny{}$0.2$}  &  {\tiny{}$>0.35$}  &  {\tiny{}$0.5$}  &  {\tiny{}$>0.35$}  &  {\tiny{}$1.0$}  &  {\tiny{}$0.000$}  &  {\tiny{}$2.0$}  &  {\tiny{}$0.000$}  &  {\tiny{}$2.168$}  &  {\tiny{}$0.000$}   & \ldots & \ldots  \tabularnewline
  \ldots  & \ldots & \ldots & \ldots  &  {\tiny{}$1.0$}  &  {\tiny{}$0.2$}  &  {\tiny{}$>0.35$}  &  {\tiny{}$0.5$}  &  {\tiny{}$>0.35$}  &  {\tiny{}$0.7$}  &  {\tiny{}$>0.35$}  &  {\tiny{}$1.0$}  &  {\tiny{}$0.000$}  &    {\tiny{}$2.0$}  &  {\tiny{}$0.000$}  &  {\tiny{}$2.168$}  &  {\tiny{}$0.000$}  \tabularnewline
  \ldots  & \ldots & \ldots & \ldots  &  {\tiny{}$10.0$}  &  {\tiny{}$0.2$}  &  {\tiny{}$>0.35$}  &  {\tiny{}$0.5$}  &  {\tiny{}$>0.35$}  &  {\tiny{}$1.0$}  &  {\tiny{}$0.014$}  &  {\tiny{}$2.0$}  &  {\tiny{}$0.006$}  &  {\tiny{}$2.168$}  &  {\tiny{}$  -  $}  & \ldots & \ldots  \tabularnewline
{\tiny{}$0.750$}  &  {\tiny{}$27.3$}  &  {\tiny{}$0.082$}  &  {\tiny{}$0.068$}  &  {\tiny{}$0.01$}  &  {\tiny{}$0.2$}  &  {\tiny{}$>0.35$}  &  {\tiny{}$0.5$}  &  {\tiny{}$>0.35$}  &  {\tiny{}$1.0$}  &  {\tiny{}$0.012$}  &  {\tiny{}$2.0$}  &  {\tiny{}$0.000$}  &  {\tiny{}$2.262$}  &  {\tiny{}$0.000$}   & \ldots & \ldots  \tabularnewline
  \ldots  & \ldots & \ldots & \ldots  &  {\tiny{}$0.1$}  &  {\tiny{}$0.2$}  &  {\tiny{}$>0.35$}  &  {\tiny{}$0.5$}  &  {\tiny{}$>0.35$}  &  {\tiny{}$1.0$}  &  {\tiny{}$0.004$}  &  {\tiny{}$2.0$}  &  {\tiny{}$0.000$}  &  {\tiny{}$2.262$}  &  {\tiny{}$0.000$}   & \ldots & \ldots  \tabularnewline
  \ldots  & \ldots & \ldots & \ldots  &  {\tiny{}$1.0$}  &  {\tiny{}$0.2$}  &  {\tiny{}$>0.35$}  &  {\tiny{}$0.5$}  &  {\tiny{}$>0.35$}  &  {\tiny{}$0.7$}  &  {\tiny{}$>0.35$}  &  {\tiny{}$1.0$}  &  {\tiny{}$0.006$}  &    {\tiny{}$2.0$}  &  {\tiny{}$0.000$}  &  {\tiny{}$2.262$}  &  {\tiny{}$0.000$}  \tabularnewline
  \ldots  & \ldots & \ldots & \ldots  &  {\tiny{}$10.0$}  &  {\tiny{}$0.2$}  &  {\tiny{}$>0.35$}  &  {\tiny{}$0.5$}  &  {\tiny{}$>0.35$}  &  {\tiny{}$1.0$}  &  {\tiny{}$0.009$}  &  {\tiny{}$2.0$}  &  {\tiny{}$0.004$}  &  {\tiny{}$2.262$}  &  {\tiny{}$  -  $}  & \ldots & \ldots  \tabularnewline
{\tiny{}$0.775$}  &  {\tiny{}$23.6$}  &  {\tiny{}$0.075$}  &  {\tiny{}$0.063$}  &  {\tiny{}$0.01$}  &  {\tiny{}$0.2$}  &  {\tiny{}$>0.35$}  &  {\tiny{}$0.5$}  &  {\tiny{}$>0.35$}  &  {\tiny{}$1.0$}  &  {\tiny{}$0.015$}  &  {\tiny{}$2.0$}  &  {\tiny{}$0.000$}  &  {\tiny{}$2.361$}  &  {\tiny{}$0.000$}   & \ldots & \ldots  \tabularnewline
  \ldots  & \ldots & \ldots & \ldots  &  {\tiny{}$0.1$}  &  {\tiny{}$0.2$}  &  {\tiny{}$>0.35$}  &  {\tiny{}$0.5$}  &  {\tiny{}$>0.35$}  &  {\tiny{}$1.0$}  &  {\tiny{}$0.001$}  &  {\tiny{}$2.0$}  &  {\tiny{}$0.000$}  &  {\tiny{}$2.361$}  &  {\tiny{}$0.000$}   & \ldots & \ldots  \tabularnewline
  \ldots  & \ldots & \ldots & \ldots  &  {\tiny{}$1.0$}  &  {\tiny{}$0.2$}  &  {\tiny{}$>0.35$}  &  {\tiny{}$0.5$}  &  {\tiny{}$>0.35$}  &  {\tiny{}$0.7$}  &  {\tiny{}$>0.35$}  &  {\tiny{}$1.0$}  &  {\tiny{}$0.002$}  &    {\tiny{}$2.0$}  &  {\tiny{}$0.000$}  &  {\tiny{}$2.361$}  &  {\tiny{}$0.000$}  \tabularnewline
  \ldots  & \ldots & \ldots & \ldots  &  {\tiny{}$10.0$}  &  {\tiny{}$0.2$}  &  {\tiny{}$>0.35$}  &  {\tiny{}$0.5$}  &  {\tiny{}$>0.35$}  &  {\tiny{}$1.0$}  &  {\tiny{}$0.004$}  &  {\tiny{}$2.0$}  &  {\tiny{}$0.000$}  &  {\tiny{}$2.361$}  &  {\tiny{}$  -  $}  & \ldots & \ldots  \tabularnewline
{\tiny{}$0.800$}  &  {\tiny{}$20.2$}  &  {\tiny{}$0.067$}  &  {\tiny{}$0.059$}  &  {\tiny{}$0.01$}  &  {\tiny{}$0.2$}  &  {\tiny{}$>0.35$}  &  {\tiny{}$0.5$}  &  {\tiny{}$>0.35$}  &  {\tiny{}$1.0$}  &  {\tiny{}$0.015$}  &  {\tiny{}$2.0$}  &  {\tiny{}$0.000$}  &  {\tiny{}$2.466$}  &  {\tiny{}$  -  $}   & \ldots & \ldots  \tabularnewline
  \ldots  & \ldots & \ldots & \ldots  &  {\tiny{}$0.1$}  &  {\tiny{}$0.2$}  &  {\tiny{}$>0.35$}  &  {\tiny{}$0.5$}  &  {\tiny{}$>0.35$}  &  {\tiny{}$1.0$}  &  {\tiny{}$0.001$}  &  {\tiny{}$2.0$}  &  {\tiny{}$0.000$}  &  {\tiny{}$2.466$}  &  {\tiny{}$0.000$}   & \ldots & \ldots  \tabularnewline
  \ldots  & \ldots & \ldots & \ldots  &  {\tiny{}$1.0$}  &  {\tiny{}$0.2$}  &  {\tiny{}$>0.35$}  &  {\tiny{}$0.5$}  &  {\tiny{}$>0.35$}  &  {\tiny{}$0.7$}  &  {\tiny{}$>0.35$}  &  {\tiny{}$1.0$}  &  {\tiny{}$0.035$}  &    {\tiny{}$2.0$}  &  {\tiny{}$0.000$}  &  {\tiny{}$2.466$}  &  {\tiny{}$0.000$}  \tabularnewline
  \ldots  & \ldots & \ldots & \ldots  &  {\tiny{}$10.0$}  &  {\tiny{}$0.2$}  &  {\tiny{}$>0.35$}  &  {\tiny{}$0.5$}  &  {\tiny{}$>0.35$}  &  {\tiny{}$1.0$}  &  {\tiny{}$0.001$}  &  {\tiny{}$2.0$}  &  {\tiny{}$0.000$}  &  {\tiny{}$2.466$}  &  {\tiny{}$  -  $}  & \ldots & \ldots  \tabularnewline
{\tiny{}$0.825$}  &  {\tiny{}$17.4$}  &  {\tiny{}$0.061$}  &  {\tiny{}$0.054$}  &  {\tiny{}$0.01$}  &  {\tiny{}$0.2$}  &  {\tiny{}$>0.35$}  &  {\tiny{}$0.5$}  &  {\tiny{}$>0.35$}  &  {\tiny{}$1.0$}  &  {\tiny{}$  -  $}  &  {\tiny{}$2.0$}  &  {\tiny{}$0.000$}  &  {\tiny{}$2.578$}  &  {\tiny{}$  -  $}   & \ldots & \ldots  \tabularnewline
  \ldots  & \ldots & \ldots & \ldots  &  {\tiny{}$0.1$}  &  {\tiny{}$0.2$}  &  {\tiny{}$>0.35$}  &  {\tiny{}$0.5$}  &  {\tiny{}$>0.35$}  &  {\tiny{}$1.0$}  &  {\tiny{}$0.016$}  &  {\tiny{}$2.0$}  &  {\tiny{}$0.000$}  &  {\tiny{}$2.578$}  &  {\tiny{}$  -  $}   & \ldots & \ldots  \tabularnewline
  \ldots  & \ldots & \ldots & \ldots  &  {\tiny{}$1.0$}  &  {\tiny{}$0.2$}  &  {\tiny{}$>0.35$}  &  {\tiny{}$0.5$}  &  {\tiny{}$>0.35$}  &  {\tiny{}$0.7$}  &  {\tiny{}$>0.35$}  &  {\tiny{}$1.0$}  &  {\tiny{}$0.038$}  &    {\tiny{}$2.0$}  &  {\tiny{}$0.000$}  &  {\tiny{}$2.578$}  &  {\tiny{}$0.000$}  \tabularnewline
  \ldots  & \ldots & \ldots & \ldots  &  {\tiny{}$10.0$}  &  {\tiny{}$0.2$}  &  {\tiny{}$>0.35$}  &  {\tiny{}$0.5$}  &  {\tiny{}$>0.35$}  &  {\tiny{}$1.0$}  &  {\tiny{}$>0.35$}  &  {\tiny{}$2.0$}  &  {\tiny{}$0.000$}  &  {\tiny{}$2.578$}  &  {\tiny{}$  -  $}  & \ldots & \ldots  \tabularnewline
\end{longtable}
\tablefoot{
All masses are in solar masses. If critical rotation was not reached by then, the CEMP and Ba star progenitor models were stopped when, respectively,
$0.35\ M_{\odot}$ and $2\ M_{\odot}$ of material were accreted.
A minus indicates failure of convergence.
The values of $t_\text{mt}$ are from \citet{2012ApJ...747....2L} and \citet{1998MNRAS.298..525P} for stars of $Z=10^{-4}$ and $Z=0.008$, respectively.
}
\end{longtab}

%% file: AA-2017-30746.bbl
\begin{thebibliography}{90}
\expandafter\ifx\csname natexlab\endcsname\relax\def\natexlab#1{#1}\fi

\bibitem[{{Abate} {et~al.}(2015{\natexlab{a}}){Abate}, {Pols}, {Izzard}, \&
  {Karakas}}]{2015A&A...581A..22A}
{Abate}, C., {Pols}, O.~R., {Izzard}, R.~G., \& {Karakas}, A.~I.
  2015{\natexlab{a}}, \aap, 581, A22

\bibitem[{{Abate} {et~al.}(2013){Abate}, {Pols}, {Izzard}, {Mohamed}, \& {de
  Mink}}]{2013A&A...552A..26A}
{Abate}, C., {Pols}, O.~R., {Izzard}, R.~G., {Mohamed}, S.~S., \& {de Mink},
  S.~E. 2013, \aap, 552, A26

\bibitem[{{Abate} {et~al.}(2015{\natexlab{b}}){Abate}, {Pols}, {Karakas}, \&
  {Izzard}}]{2015A&A...576A.118A}
{Abate}, C., {Pols}, O.~R., {Karakas}, A.~I., \& {Izzard}, R.~G.
  2015{\natexlab{b}}, \aap, 576, A118

\bibitem[{{Abate} {et~al.}(2015{\natexlab{c}}){Abate}, {Pols}, {Stancliffe},
  {Izzard}, {Karakas}, {Beers}, \& {Lee}}]{2015A&A...581A..62A}
{Abate}, C., {Pols}, O.~R., {Stancliffe}, R.~J., {et~al.} 2015{\natexlab{c}},
  \aap, 581, A62

\bibitem[{{Armitage} \& {Clarke}(1996)}]{1996MNRAS.280..458A}
{Armitage}, P.~J. \& {Clarke}, C.~J. 1996, \mnras, 280, 458

\bibitem[{{Beers} \& {Christlieb}(2005)}]{2005ARA&A..43..531B}
{Beers}, T.~C. \& {Christlieb}, N. 2005, \araa, 43, 531

\bibitem[{{Bidelman} \& {Keenan}(1951)}]{1951ApJ...114..473B}
{Bidelman}, W.~P. \& {Keenan}, P.~C. 1951, \apj, 114, 473

\bibitem[{{Blondin} \& {Raymer}(2012)}]{2012ApJ...752...30B}
{Blondin}, J.~M. \& {Raymer}, E. 2012, \apj, 752, 30

\bibitem[{{B{\"o}hm-Vitense} {et~al.}(2000){B{\"o}hm-Vitense}, {Carpenter},
  {Robinson}, {Ake}, \& {Brown}}]{2000ApJ...533..969B}
{B{\"o}hm-Vitense}, E., {Carpenter}, K., {Robinson}, R., {Ake}, T., \& {Brown},
  J. 2000, \apj, 533, 969

\bibitem[{{Bond}(1974)}]{1974ApJ...194...95B}
{Bond}, H.~E. 1974, \apj, 194, 95

\bibitem[{{Bond} {et~al.}(2003){Bond}, {Pollacco}, \&
  {Webbink}}]{2003AJ....125..260B}
{Bond}, H.~E., {Pollacco}, D.~L., \& {Webbink}, R.~F. 2003, \aj, 125, 260

\bibitem[{{Bondi} \& {Hoyle}(1944)}]{1944MNRAS.104..273B}
{Bondi}, H. \& {Hoyle}, F. 1944, \mnras, 104, 273

\bibitem[{{Chen} {et~al.}(2017){Chen}, {Frank}, {Blackman}, {Nordhaus}, \&
  {Carroll-Nellenback}}]{2017MNRAS.468.4465C}
{Chen}, Z., {Frank}, A., {Blackman}, E.~G., {Nordhaus}, J., \&
  {Carroll-Nellenback}, J. 2017, \mnras, 468, 4465

\bibitem[{{de Val-Borro} {et~al.}(2009){de Val-Borro}, {Karovska}, \&
  {Sasselov}}]{2009ApJ...700.1148D}
{de Val-Borro}, M., {Karovska}, M., \& {Sasselov}, D. 2009, \apj, 700, 1148

\bibitem[{{Dervi{\c s}o{\v g}lu} {et~al.}(2010){Dervi{\c s}o{\v g}lu}, {Tout},
  \& {Ibano{\v g}lu}}]{2010MNRAS.406.1071D}
{Dervi{\c s}o{\v g}lu}, A., {Tout}, C.~A., \& {Ibano{\v g}lu}, C. 2010, \mnras,
  406, 1071

\bibitem[{{Edgar}(2004)}]{2004NewAR..48..843E}
{Edgar}, R. 2004, \nar, 48, 843

\bibitem[{{Eggleton}(2006)}]{2006epbm.book.....E}
{Eggleton}, P. 2006, {Evolutionary Processes in Binary and Multiple Stars}

\bibitem[{{Eggleton}(1971)}]{1971MNRAS.151..351E}
{Eggleton}, P.~P. 1971, \mnras, 151, 351

\bibitem[{{Eggleton}(1972)}]{1972MNRAS.156..361E}
{Eggleton}, P.~P. 1972, \mnras, 156, 361

\bibitem[{{Endal} \& {Sofia}(1976)}]{1976ApJ...210..184E}
{Endal}, A.~S. \& {Sofia}, S. 1976, \apj, 210, 184

\bibitem[{{Endal} \& {Sofia}(1978)}]{1978ApJ...220..279E}
{Endal}, A.~S. \& {Sofia}, S. 1978, \apj, 220, 279

\bibitem[{{Fricke}(1968)}]{1968ZA.....68..317F}
{Fricke}, K. 1968, \zap, 68, 317

\bibitem[{{Georgy} {et~al.}(2011){Georgy}, {Meynet}, \&
  {Maeder}}]{2011A&A...527A..52G}
{Georgy}, C., {Meynet}, G., \& {Maeder}, A. 2011, \aap, 527, A52

\bibitem[{{Goldreich} \& {Schubert}(1967)}]{1967ApJ...150..571G}
{Goldreich}, P. \& {Schubert}, G. 1967, \apj, 150, 571

\bibitem[{{Haemmerl{\'e}} {et~al.}(2016){Haemmerl{\'e}}, {Eggenberger},
  {Meynet}, {Maeder}, \& {Charbonnel}}]{2016A&A...585A..65H}
{Haemmerl{\'e}}, L., {Eggenberger}, P., {Meynet}, G., {Maeder}, A., \&
  {Charbonnel}, C. 2016, \aap, 585, A65

\bibitem[{{Hameury} \& {Lasota}(2005)}]{2005A&A...443..283H}
{Hameury}, J.-M. \& {Lasota}, J.-P. 2005, \aap, 443, 283

\bibitem[{{Han} {et~al.}(1995){Han}, {Eggleton}, {Podsiadlowski}, \&
  {Tout}}]{1995MNRAS.277.1443H}
{Han}, Z., {Eggleton}, P.~P., {Podsiadlowski}, P., \& {Tout}, C.~A. 1995,
  \mnras, 277, 1443

\bibitem[{{Hansen} {et~al.}(2016){Hansen}, {Andersen}, {Nordstr{\"o}m},
  {Beers}, {Placco}, {Yoon}, \& {Buchhave}}]{2016A&A...588A...3H}
{Hansen}, T.~T., {Andersen}, J., {Nordstr{\"o}m}, B., {et~al.} 2016, \aap, 588,
  A3

\bibitem[{{Hartmann} {et~al.}(1998){Hartmann}, {Calvet}, {Gullbring}, \&
  {D'Alessio}}]{1998ApJ...495..385H}
{Hartmann}, L., {Calvet}, N., {Gullbring}, E., \& {D'Alessio}, P. 1998, \apj,
  495, 385

\bibitem[{{Heger} {et~al.}(2000){Heger}, {Langer}, \&
  {Woosley}}]{2000ApJ...528..368H}
{Heger}, A., {Langer}, N., \& {Woosley}, S.~E. 2000, \apj, 528, 368

\bibitem[{{Hosokawa} {et~al.}(2010){Hosokawa}, {Yorke}, \&
  {Omukai}}]{2010ApJ...721..478H}
{Hosokawa}, T., {Yorke}, H.~W., \& {Omukai}, K. 2010, \apj, 721, 478

\bibitem[{{Huarte-Espinosa} {et~al.}(2013){Huarte-Espinosa},
  {Carroll-Nellenback}, {Nordhaus}, {Frank}, \&
  {Blackman}}]{2013MNRAS.433..295H}
{Huarte-Espinosa}, M., {Carroll-Nellenback}, J., {Nordhaus}, J., {Frank}, A.,
  \& {Blackman}, E.~G. 2013, \mnras, 433, 295

\bibitem[{{Ichikawa} \& {Osaki}(1994)}]{1994PASJ...46..621I}
{Ichikawa}, S. \& {Osaki}, Y. 1994, \pasj, 46, 621

\bibitem[{{Izzard} {et~al.}(2010){Izzard}, {Dermine}, \&
  {Church}}]{2010A&A...523A..10I}
{Izzard}, R.~G., {Dermine}, T., \& {Church}, R.~P. 2010, \aap, 523, A10

\bibitem[{{Jeffries} \& {Smalley}(1996)}]{1996A&A...315L..19J}
{Jeffries}, R.~D. \& {Smalley}, B. 1996, \aap, 315, L19

\bibitem[{{Jorissen} {et~al.}(2016){Jorissen}, {Van Eck}, {Van Winckel},
  {Merle}, {Boffin}, {Andersen}, {Nordstr{\"o}m}, {Udry}, {Masseron},
  {Lenaerts}, \& {Waelkens}}]{2016A&A...586A.158J}
{Jorissen}, A., {Van Eck}, S., {Van Winckel}, H., {et~al.} 2016, \aap, 586,
  A158

\bibitem[{{Karakas}(2010)}]{2010MNRAS.403.1413K}
{Karakas}, A.~I. 2010, \mnras, 403, 1413

\bibitem[{{Keenan}(1942)}]{1942ApJ....96..101K}
{Keenan}, P.~C. 1942, \apj, 96, 101

\bibitem[{{Kellett} {et~al.}(1995){Kellett}, {Bromage}, {Brown}, {Jeffries},
  {James}, {Kilkenny}, {Robb}, {Wonnacott}, {Lloyd}, \&
  {Clayton}}]{1995ApJ...438..364K}
{Kellett}, B.~J., {Bromage}, G.~E., {Brown}, A., {et~al.} 1995, \apj, 438, 364

\bibitem[{{Kippenhahn}(1974)}]{1974IAUS...66...20K}
{Kippenhahn}, R. 1974, in IAU Symposium, Vol.~66, Late Stages of Stellar
  Evolution, ed. R.~J. {Tayler} \& J.~E. {Hesser}, 20

\bibitem[{{Krti{\v c}ka} {et~al.}(2011){Krti{\v c}ka}, {Owocki}, \&
  {Meynet}}]{2011A&A...527A..84K}
{Krti{\v c}ka}, J., {Owocki}, S.~P., \& {Meynet}, G. 2011, \aap, 527, A84

\bibitem[{{Lee} {et~al.}(2013){Lee}, {Beers}, {Masseron}, {Plez}, {Rockosi},
  {Sobeck}, {Yanny}, {Lucatello}, {Sivarani}, {Placco}, \&
  {Carollo}}]{2013AJ....146..132L}
{Lee}, Y.~S., {Beers}, T.~C., {Masseron}, T., {et~al.} 2013, \aj, 146, 132

\bibitem[{{Lin} \& {Pringle}(1976)}]{1976IAUS...73..237L}
{Lin}, D.~N.~C. \& {Pringle}, J.~E. 1976, in IAU Symposium, Vol.~73, Structure
  and Evolution of Close Binary Systems, ed. P.~{Eggleton}, S.~{Mitton}, \&
  J.~{Whelan}, 237

\bibitem[{{Lodato}(2008)}]{2008NewAR..52...21L}
{Lodato}, G. 2008, \nar, 52, 21

\bibitem[{{Lucatello} {et~al.}(2006){Lucatello}, {Beers}, {Christlieb},
  {Barklem}, {Rossi}, {Marsteller}, {Sivarani}, \& {Lee}}]{2006ApJ...652L..37L}
{Lucatello}, S., {Beers}, T.~C., {Christlieb}, N., {et~al.} 2006, \apjl, 652,
  L37

\bibitem[{{Lucatello} \& {Gratton}(2003)}]{2003A&A...406..691L}
{Lucatello}, S. \& {Gratton}, R.~G. 2003, \aap, 406, 691

\bibitem[{{Lugaro} {et~al.}(2012){Lugaro}, {Karakas}, {Stancliffe}, \&
  {Rijs}}]{2012ApJ...747....2L}
{Lugaro}, M., {Karakas}, A.~I., {Stancliffe}, R.~J., \& {Rijs}, C. 2012, \apj,
  747, 2

\bibitem[{{Maeder}(2009)}]{2009pfer.book.....M}
{Maeder}, A. 2009, {Physics, Formation and Evolution of Rotating Stars}

\bibitem[{{Marigo} \& {Girardi}(2007)}]{2007A&A...469..239M}
{Marigo}, P. \& {Girardi}, L. 2007, \aap, 469, 239

\bibitem[{{Masseron} {et~al.}(2012){Masseron}, {Johnson}, {Lucatello},
  {Karakas}, {Plez}, {Beers}, \& {Christlieb}}]{2012ApJ...751...14M}
{Masseron}, T., {Johnson}, J.~A., {Lucatello}, S., {et~al.} 2012, \apj, 751, 14

\bibitem[{{Mastrodemos} \& {Morris}(1998)}]{1998ApJ...497..303M}
{Mastrodemos}, N. \& {Morris}, M. 1998, \apj, 497, 303

\bibitem[{{Matrozis} \& {Stancliffe}(2016)}]{2016A&A...592A..29M}
{Matrozis}, E. \& {Stancliffe}, R.~J. 2016, \aap, 592, A29

\bibitem[{{Matrozis} \& {Stancliffe}(in press)}]{Matrozis2017inpress}
{Matrozis}, E. \& {Stancliffe}, R.~J., \aap, in press

\bibitem[{{Matt} \& {Pudritz}(2005{\natexlab{a}})}]{2005ApJ...632L.135M}
{Matt}, S. \& {Pudritz}, R.~E. 2005{\natexlab{a}}, \apjl, 632, L135

\bibitem[{{Matt} \& {Pudritz}(2005{\natexlab{b}})}]{2005MNRAS.356..167M}
{Matt}, S. \& {Pudritz}, R.~E. 2005{\natexlab{b}}, \mnras, 356, 167

\bibitem[{{McClure} \& {Woodsworth}(1990)}]{1990ApJ...352..709M}
{McClure}, R.~D. \& {Woodsworth}, A.~W. 1990, \apj, 352, 709

\bibitem[{{Merle} {et~al.}(2016){Merle}, {Jorissen}, {Van Eck}, {Masseron}, \&
  {Van Winckel}}]{2016A&A...586A.151M}
{Merle}, T., {Jorissen}, A., {Van Eck}, S., {Masseron}, T., \& {Van Winckel},
  H. 2016, \aap, 586, A151

\bibitem[{{Meynet} {et~al.}(2006){Meynet}, {Ekstr{\"o}m}, \&
  {Maeder}}]{2006A&A...447..623M}
{Meynet}, G., {Ekstr{\"o}m}, S., \& {Maeder}, A. 2006, \aap, 447, 623

\bibitem[{{Meynet} \& {Maeder}(1997)}]{1997A&A...321..465M}
{Meynet}, G. \& {Maeder}, A. 1997, \aap, 321, 465

\bibitem[{{Miszalski} {et~al.}(2012){Miszalski}, {Boffin}, {Frew}, {Acker},
  {K{\"o}ppen}, {Moffat}, \& {Parker}}]{2012MNRAS.419...39M}
{Miszalski}, B., {Boffin}, H.~M.~J., {Frew}, D.~J., {et~al.} 2012, \mnras, 419,
  39

\bibitem[{{Miszalski} {et~al.}(2013){Miszalski}, {Boffin}, {Jones}, {Karakas},
  {K{\"o}ppen}, {Tyndall}, {Mohamed}, {Rodr{\'{\i}}guez-Gil}, \&
  {Santander-Garc{\'{\i}}a}}]{2013MNRAS.436.3068M}
{Miszalski}, B., {Boffin}, H.~M.~J., {Jones}, D., {et~al.} 2013, \mnras, 436,
  3068

\bibitem[{{Mohamed} \& {Podsiadlowski}(2007)}]{2007ASPC..372..397M}
{Mohamed}, S. \& {Podsiadlowski}, P. 2007, in Astronomical Society of the
  Pacific Conference Series, Vol. 372, 15th European Workshop on White Dwarfs,
  ed. R.~{Napiwotzki} \& M.~R. {Burleigh}, 397

\bibitem[{{North} \& {Duquennoy}(1991)}]{1991A&A...244..335N}
{North}, P. \& {Duquennoy}, A. 1991, \aap, 244, 335

\bibitem[{{Packet}(1981)}]{1981A&A...102...17P}
{Packet}, W. 1981, \aap, 102, 17

\bibitem[{{Papaloizou} \& {Pringle}(1977)}]{1977MNRAS.181..441P}
{Papaloizou}, J. \& {Pringle}, J.~E. 1977, \mnras, 181, 441

\bibitem[{{Perets} \& {Kenyon}(2013)}]{2013ApJ...764..169P}
{Perets}, H.~B. \& {Kenyon}, S.~J. 2013, \apj, 764, 169

\bibitem[{{Placco} {et~al.}(2014){Placco}, {Frebel}, {Beers}, \&
  {Stancliffe}}]{2014ApJ...797...21P}
{Placco}, V.~M., {Frebel}, A., {Beers}, T.~C., \& {Stancliffe}, R.~J. 2014,
  \apj, 797, 21

\bibitem[{{Pols} {et~al.}(2003){Pols}, {Karakas}, {Lattanzio}, \&
  {Tout}}]{2003ASPC..303..290P}
{Pols}, O.~R., {Karakas}, A.~I., {Lattanzio}, J.~C., \& {Tout}, C.~A. 2003, in
  Astronomical Society of the Pacific Conference Series, Vol. 303, Symbiotic
  Stars Probing Stellar Evolution, ed. R.~L.~M. {Corradi}, J.~{Mikolajewska},
  \& T.~J. {Mahoney}, 290

\bibitem[{{Pols} {et~al.}(1998){Pols}, {Schr{\"o}der}, {Hurley}, {Tout}, \&
  {Eggleton}}]{1998MNRAS.298..525P}
{Pols}, O.~R., {Schr{\"o}der}, K.-P., {Hurley}, J.~R., {Tout}, C.~A., \&
  {Eggleton}, P.~P. 1998, \mnras, 298, 525

\bibitem[{{Pols} {et~al.}(1995){Pols}, {Tout}, {Eggleton}, \&
  {Han}}]{1995MNRAS.274..964P}
{Pols}, O.~R., {Tout}, C.~A., {Eggleton}, P.~P., \& {Han}, Z. 1995, \mnras,
  274, 964

\bibitem[{{Popham} \& {Narayan}(1991)}]{1991ApJ...370..604P}
{Popham}, R. \& {Narayan}, R. 1991, \apj, 370, 604

\bibitem[{{Potter} {et~al.}(2012{\natexlab{a}}){Potter}, {Tout}, \&
  {Brott}}]{2012MNRAS.423.1221P}
{Potter}, A.~T., {Tout}, C.~A., \& {Brott}, I. 2012{\natexlab{a}}, \mnras, 423,
  1221

\bibitem[{{Potter} {et~al.}(2012{\natexlab{b}}){Potter}, {Tout}, \&
  {Eldridge}}]{2012MNRAS.419..748P}
{Potter}, A.~T., {Tout}, C.~A., \& {Eldridge}, J.~J. 2012{\natexlab{b}},
  \mnras, 419, 748

\bibitem[{{Prialnik} \& {Livio}(1985)}]{1985MNRAS.216...37P}
{Prialnik}, D. \& {Livio}, M. 1985, \mnras, 216, 37

\bibitem[{{Pringle}(1981)}]{1981ARA&A..19..137P}
{Pringle}, J.~E. 1981, \araa, 19, 137

\bibitem[{{Ramstedt} \& {Olofsson}(2014)}]{2014A&A...566A.145R}
{Ramstedt}, S. \& {Olofsson}, H. 2014, \aap, 566, A145

\bibitem[{{Sarna} \& {Ziolkowski}(1988)}]{1988AcA....38...89S}
{Sarna}, M.~J. \& {Ziolkowski}, J. 1988, \actaa, 38, 89

\bibitem[{{Schwarzenberg-Czerny} \& {Rozyczka}(1988)}]{1988AcA....38..189S}
{Schwarzenberg-Czerny}, A. \& {Rozyczka}, M. 1988, \actaa, 38, 189

\bibitem[{{Soker} \& {Rappaport}(2000)}]{2000ApJ...538..241S}
{Soker}, N. \& {Rappaport}, S. 2000, \apj, 538, 241

\bibitem[{{Stancliffe} \& {Eldridge}(2009)}]{2009MNRAS.396.1699S}
{Stancliffe}, R.~J. \& {Eldridge}, J.~J. 2009, \mnras, 396, 1699

\bibitem[{{Stancliffe} \& {Glebbeek}(2008)}]{2008MNRAS.389.1828S}
{Stancliffe}, R.~J. \& {Glebbeek}, E. 2008, \mnras, 389, 1828

\bibitem[{{Starkenburg} {et~al.}(2014){Starkenburg}, {Shetrone}, {McConnachie},
  \& {Venn}}]{2014MNRAS.441.1217S}
{Starkenburg}, E., {Shetrone}, M.~D., {McConnachie}, A.~W., \& {Venn}, K.~A.
  2014, \mnras, 441, 1217

\bibitem[{{Talon} \& {Zahn}(1997)}]{1997A&A...317..749T}
{Talon}, S. \& {Zahn}, J.-P. 1997, \aap, 317, 749

\bibitem[{{Theuns} {et~al.}(1996){Theuns}, {Boffin}, \&
  {Jorissen}}]{1996MNRAS.280.1264T}
{Theuns}, T., {Boffin}, H.~M.~J., \& {Jorissen}, A. 1996, \mnras, 280, 1264

\bibitem[{{van Loon} {et~al.}(2005){van Loon}, {Cioni}, {Zijlstra}, \&
  {Loup}}]{2005A&A...438..273V}
{van Loon}, J.~T., {Cioni}, M.-R.~L., {Zijlstra}, A.~A., \& {Loup}, C. 2005,
  \aap, 438, 273

\bibitem[{{Vassiliadis} \& {Wood}(1993)}]{1993ApJ...413..641V}
{Vassiliadis}, E. \& {Wood}, P.~R. 1993, \apj, 413, 641

\bibitem[{{Vennes} {et~al.}(1998){Vennes}, {Christian}, \&
  {Thorstensen}}]{1998ApJ...502..763V}
{Vennes}, S., {Christian}, D.~J., \& {Thorstensen}, J.~R. 1998, \apj, 502, 763

\bibitem[{{Wasiutynski}(1946)}]{1946ApNr....4....1W}
{Wasiutynski}, J. 1946, Astrophysica Norvegica, 4, 1

\bibitem[{{Zahn}(1974)}]{1974IAUS...59..185Z}
{Zahn}, J.-P. 1974, in IAU Symposium, Vol.~59, Stellar Instability and
  Evolution, ed. P.~{Ledoux}, A.~{Noels}, \& A.~W. {Rodgers}, 185--194

\bibitem[{{Zahn}(1977)}]{1977A&A....57..383Z}
{Zahn}, J.-P. 1977, \aap, 57, 383

\bibitem[{{Zahn}(1992)}]{1992A&A...265..115Z}
{Zahn}, J.-P. 1992, \aap, 265, 115

\end{thebibliography}
